\documentclass[aps,prb,twocolumn,superscriptaddress,showpacs]{revtex4-2}
\usepackage{amsmath,amssymb,wasysym,graphicx}

\usepackage[utf8]{inputenc}
\usepackage[T1]{fontenc}
\usepackage{xcolor}

\IfFileExists{newtxtext.sty}
   {\usepackage{newtxtext,newtxmath}}
   {\IfFileExists{stix.sty}
      {\usepackage{stix}}
      {\IfFileExists{mathptmx.sty}
      {\usepackage{mathptmx}}{} } }

\usepackage{textcomp}

\usepackage{bm}

\IfFileExists{siunitx.sty}{\usepackage{booktabs,siunitx}}{}

\usepackage{color}
\definecolor{LinkColor}{rgb}{0.256,0.439,0.588}
\usepackage{hyperref}
\hypersetup{
   pdfauthor={good guys},
   pdftitle={good title},
   colorlinks=true,
   citecolor=LinkColor,
   linkcolor=LinkColor,
   urlcolor=LinkColor
}

\usepackage{pifont}
\usepackage[normalem]{ulem}

\newcommand{\RNum}[1]{\uppercase\expandafter{\romannumeral #1\relax}}

\usepackage{multirow}

\begin{document}

\title{Phase Diagram of Triangular Lattice Quantum Ising Model under External Field}

\author{Yuan Da Liao}\thanks{These two authors contributed equally}
\affiliation{Beijing National Laboratory for Condensed Matter Physics and Institute of Physics, Chinese Academy of Sciences, Beijing 100190, China}
\affiliation{School of Physical Sciences, University of Chinese Academy of Sciences, Beijing 100190, China}

\author{Han Li}\thanks{These two authors contributed equally}
\affiliation{School of Physics, Beihang University, Beijing 100191, China}

\author{Zheng Yan}
\affiliation{Department of Physics and HKU-UCAS Joint Institute of Theoretical and Computational Physics, The University of Hong Kong, Pokfulam Road, Hong Kong SAR, China}
\affiliation{State Key Laboratory of Surface Physics, Fudan University, Shanghai 200438, China}

\author{Hao-Tian Wei}
\affiliation{State Key Laboratory of Surface Physics, Fudan University, Shanghai 200438, China}
\affiliation{Center for Field Theory and Particle Physics, Department of Physics, Fudan University, Shanghai 200433, China}

\author{Wei Li}
\email{w.li@buaa.edu.cn}
\affiliation{School of Physics, Beihang University, Beijing 100191, China}
\affiliation{International Research Institute of Multidisciplinary Science, Beihang University, Beijing 100191, China}

\author{Yang Qi}
\email{qiyang@fudan.edu.cn}
\affiliation{State Key Laboratory of Surface Physics, Fudan University, Shanghai 200438, China}
\affiliation{Center for Field Theory and Particle Physics, Department of Physics, Fudan University, Shanghai 200433, China}

\author{Zi Yang Meng}
\email{zymeng@hku.hk}
\affiliation{Department of Physics and HKU-UCAS Joint Institute of Theoretical and Computational Physics, The University of Hong Kong, Pokfulam Road, Hong Kong SAR, China}

\date{\today}

\begin{abstract}
Quantum Ising model on a triangular lattice hosts a finite temperature
Berezinskii-Kosterlitz-Thouless (BKT) phase with emergent U(1) symmetry,
and it will transit into an up-up-down (UUD) phase with $C_3$ symmetry breaking
upon an infinitesimal external field along the longitudinal direction,
but the overall phase diagram spanned by the axes of external field and temperature
remains opaque due to the lack of systematic invesitgations with controlled methodologies.
By means of quantum Monte Carlo at finite temperature and ground state density matrix
renormalization group simulations, we map out the phase diagram of triangular quantum Ising model.
Starting from the upper BKT temperature at zero field, we obtain the phase boundary between the
UUD and paramagnetic phases with its 2D $q=3$ Potts universality at weak field
and weakly first order transition at strong field. Originated from the lower BKT temperature
at zero field, we analyze the low temperature phase boundary between the clock phase and the UUD phase with Ising symmetry breaking at weak fields and the quantum phase transition between the UUD and fully polarized phases at strong fields.
The accurate many-body numerical results are consistent with our field theoretical analysis.
The experimental relevance towards the BKT magnet TmMgGaO$_4$ and
programmable quantum simulators are also discussed.
\end{abstract}

\maketitle

\section{Introduction}
Frustrated magnets can host
intriguing quantum many-body states and phenomena
that trigger great research interest recently.
For example, the rare-earth compound TmMgGaO$_4$ (TMGO)
~\cite{Cava2018,Shen2019,Li2020,ZHu2020} is revealed to realize
a triangular lattice quantum Ising (TLI) model exhibiting the
Berezinskii-Kosterlitz-Thouless (BKT) phase transitions~\cite{Lih2020,Liu2020Intrinsic,ZHu2020}.
The Ising spins in the TLI model cooperate in the way such that there
emerges an U(1) symmetry at finite temperature in the BKT phase,
with power-law spin correlation and separating the low-$T$
clock ordered and high-$T$ paramagnetic phases with two
BKT phase transitioins
\cite{Moessner2001,Isakov2003,Lih2020,YCWang2017}.

Recently, measurements have been performed to explore
the BKT physics in this materials~\cite{ZHu2020,Dun2020neutron}.
In particular, the nuclear magnetic resonance (NMR) probe
found an extended regime with strong fluctuations that is
consistent with an intermediate BKT phase~\cite{ZHu2020},
and the inelastic neutron scattering (INS)
detection also suggest the existence of vortex characteristics
at finite temperature~\cite{Dun2020neutron}.
Under external fields $h$, there is also strong interest and
experimental progress on the thermodynamics and dynamical
properties of the material~\cite{Qin2020fieldtuned,Liu2020Intrinsic,Huang2020emergent},
which raised questions and inspire theoretical effort to
understand the $h$-$T$ phase diagram of the TLI model.
Despite some initial efforts, there remains open questions on the nature of field-induced ordered phases and the
universality class of phase transitions between such ordered
and the paramagnetic phase above the upper BKT temperature
and the clock ordered phase below, which are very relevant
for the understanding of TMGO experiments.

\begin{figure}[htb]
\includegraphics[width=0.9\columnwidth]{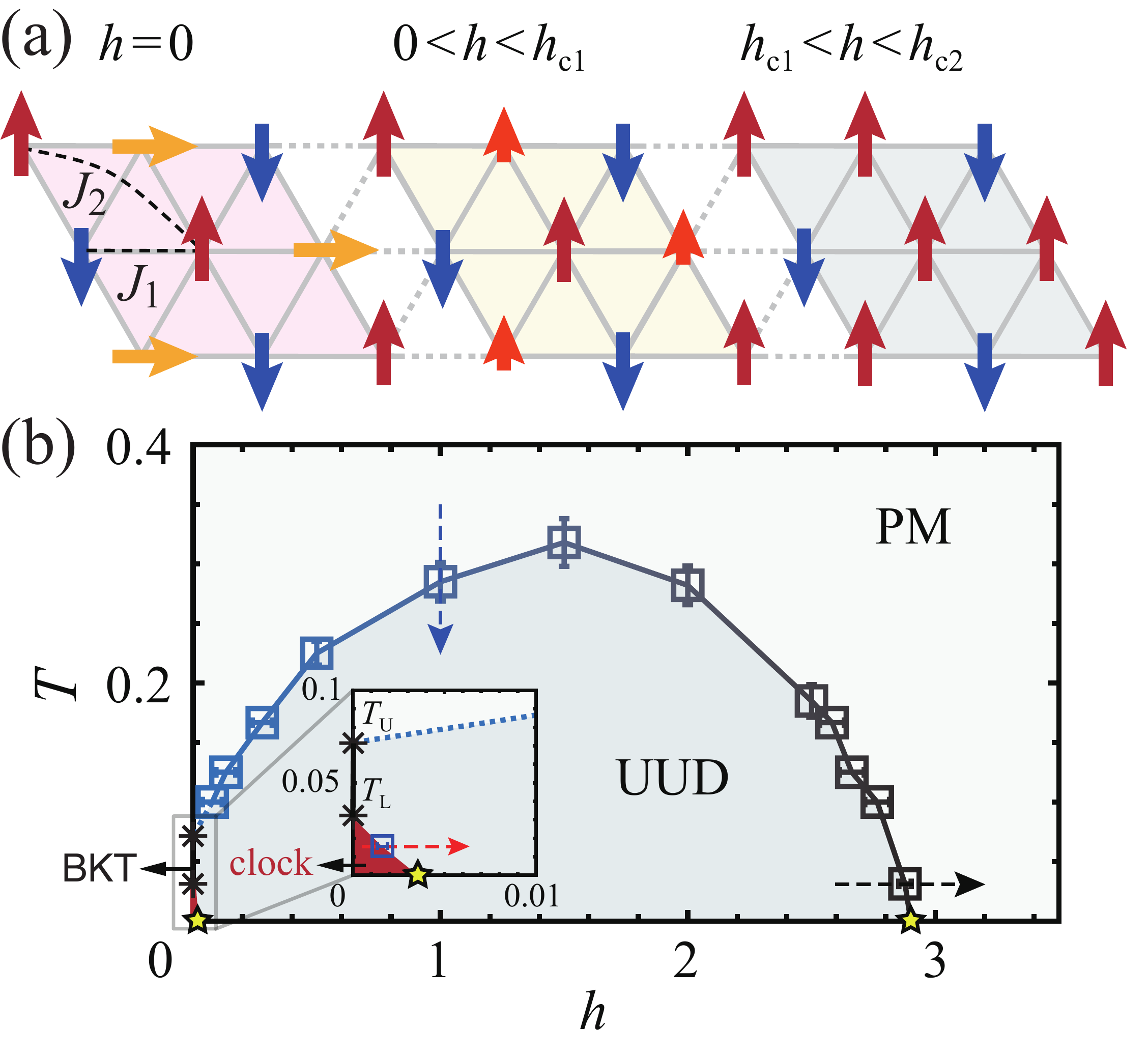}
\caption{(a) Phases at low temperature,
starting from the clock phase at zero external magnetic field (left)
to the slightly polarized clock phase at weak field (middle),
and the UUD phase under intermediate field (right).
The spin-up and spin-down arrows are all along the
external longitudinal field $h$ and the superposition
$\mid \to \rangle$ is along the intrinsic transverse field $\Delta$. 
(b) The $h$-$T$ phase diagram of the model in Eq.~\eqref{Eq:qTLI}
with $J_1=1$, $J_2=0$ and $\Delta=0.2$. The two asterisks ($\ast$ at $T_U$ and $T_L)$
along the $T$-axis stand for the upper and lower BKT temperatures
of the model at $h=0$. The blue phase boundary is the continuous
phase transition with 2D $q=3$ Potts universality starting from the $T_U$,
which separates the PM from the UUD phases;
the black phase boundary is the weakly first-order transition line
separating the same two phases. Inset zooms in at small $h$ and $T$
with the BKT phase at $h=0$ as well as the clock phase (red area)
and its Ising transition to the UUD phase depicted.
The Ising transition originates from the $T_L$ at $h=0$.
The precise position of the phase boundaries are obtained
from QMC (open squares) and DMRG (yellow stars),
and the shape of the phase diagram is consistent with
the RG analysis(see discussions in Appendices~\ref{app:RG}). The red, blue, and black
dashed arrows indicate the representative parameter paths
used in Figs.~\ref{fig:fig2} and ~\ref{fig:fig3}, respectively.}
\label{fig:fig1}
\end{figure}

In this work, we combine multiple quantum many-body theoretical tools,
including finite-temperature quantum Monte Carlo (QMC),
ground-state density matrix renormalization group (DMRG)
and field-theoretical renormalization group (RG) analysis,
to perform an unbiased study of the TLI model
under the longitudinal fields. At the ground state, we found two field-induced quantum phase
transitions (QPTs) that a lower-field
Ising QPT (at $h_{c1}$) separating the clock and up-up-down (UUD)
phases and a weakly first-order QPT (at $h_{c2}$) between
the UUD and polarized phases. We also studied the finite-temperature
fate of these two QPTs: The lower QPT field $h_{c1}$
reduces as $T$ increases, and the thermal phase transition
of 2D Ising type is continuously connected to the lower
BKT transition at zero field; on the other hand,
the first-order QPT at $h_{c2}$ also persists at finite $T$,
with transition field also reduces as $T$ increases and turns into a continuous
thermal phase transition of 2D $q=3$ Potts universality through
a tricritical point, and eventually connects to the higher
BKT transition at zero field (cf. the phase diagram in
Fig.~\ref{fig:fig1} (b)), unified with the previous understanding
of the zero-field phase diagram of the TLI model~\cite{Isakov2003,YCWang2017,Lih2020}.
Such a phase diagram is also consistent with the experimental
observations at hand and can be used to guide future ones.

\section{Model, methods, and order parameter}
We study the TLI model under the external longitudinal field as
\begin{equation}
H = J_1\sum_{\langle i,j \rangle} S_i^z S_j^z + J_2 \sum_{\langle\langle i,j \rangle\rangle} S_i^z S_j^z - \sum_i (\Delta S_i^x + h S_i^z),
\label{Eq:qTLI}
\end{equation}
where $J_1=1$ is the nearest-neighbor antiferrimagnetic interaction and set as the energy unit, and $J_2$ represents a small next-nearest neighbor interaction. Although setting $J_2=0$ in this work, we note that the small $J_2$ in the model and material could trigger interesting incommensurate phase and string magnetic excitations as discussed in Refs.~\cite{ZhengZhou202005,ZhengZhou202010}. $\Delta = 0.2$ represents the intrinsic transverse field in the material and $h$ represents the external magnetic field along longitudinal direction. Fig.~\ref{fig:fig1}(a) is the sketch of the 2D triangular lattice, with the $J_1$, $J_2$ interactions and the spin orientation from the clock phase (left, $h=0$) to the slightly polarized clock phase at intermediate field (middle, $0<h<h_{c1}$), and the UUD phase (right, $h_{c1}<h<h_{c2}$). The obtained phase diagram is shown in Fig.~\ref{fig:fig1}(b), where the rich phases and transitions will be discussed below.

To solve the model in an unbiased manner, we implement the stochastic series expansion QMC method~\cite{Sandvik2010,sandvikTFIM,melko2013stochastic,JRZhao2020} to simulate the model on finite lattice sizes and temperatures (with the linear system size $L$ upto 21 and $T$ down to $1/64$), and by means of the finite size analysis, map out the finite-temperature phase diagram and distinguish the different phase transitions. The ground state properties are computed with DMRG on a cylindrical geometry of size YC $W \times L$ with $W$ up to 12 and $L$ to 45. The retained states are up to $D=1024$, with truncation errors $\epsilon \lesssim 10^{-7}$ guaranteeing the convergence of the DMRG calculations. The phase diagram has also been analyzed with RG in qualitative manner. Details of the numerical and field-theoretical implementations are given in Appendices~\ref{app:RG} and ~\ref{app:DMRG}.

The TLI model without external field has been thoroughly studied~\cite{Moessner2001,Isakov2003,Lih2020,YCWang2017}. As shown in Fig.~\ref{fig:fig1}(b), along the $T$-axis, starting from the paramagnetic phase (PM) the system will first enter a BKT phase with emergent U(1) symmetry and power-law spin correlations and then turn into a clock phase at lower temperature. The upper and lower BKT temperatures are
denoted as the asterisks ($\ast$) in Fig.~\ref{fig:fig1}(b). The clock phase (and the UUD phase under external field) can be conveniently detected by the following complex order parameter~\cite{Isakov2003,YCWang2017}, $me^{i\theta} \equiv (m_1 + m_2 e^{i (4\pi /3)} + m_3 e^{i (-4\pi/3)})/\sqrt{3}$, where the $m_i$ with $i=1,2,3$ represent the magnetization of the three sublattices;
$m$ and $\theta$ represent the magnitude and the phase of the complex order parameter, respectively.
$me^{i\theta}$ exhibits an emergent U(1) symmmetry in the BKT phase,
and falls into one of six minima at $(2n-1)\frac\pi6$ in the clock phase,
with $n=1,2,\cdots,6$ [cf. the red dots in Fig.~\ref{fig:fig2}(f)].

When an external field $h$ is applied, the model develops an UUD phase. RG analysis in Appendices~\ref{app:RG} shows that $h$ is a relevant perturbation in the BKT phase, leading to the UUD phase at infinitesimal field.
However, above and below the BKT temperature range, the PM and clock phases are both gapped, and thus stable under an infinitesimal field.
Therefore, it is expected that the phase boundaries between the UUD phase and the PM/clock phases appear at finite $h$, extrapolating to zero field at the upper(lower) temperature of the BKT phase, respectively.

The universality class of the phase transitions can be understood as follows.
The UUD phase can also be represented by the complex order parameter $me^{i\theta}$ with $\theta$ taking one of three minima at $\pi$ and $\pm\pi/3$ [the green dots in Fig.~\ref{fig:fig2}(f)].
On the other hand, the clock phase, in the presence of $h$, still has six minima.
However, their locations move to $(2n-1)\frac\pi6+(-1)^{n-1}\delta$, where $\delta$ is an $h$-dependent angle, as indicated by the arrows in Fig.~\ref{fig:fig2}(f).
When $\delta=0$, the minima are evenly distributed around a circle in the complex plane, which represent the clock state at $h=0$~\cite{YCWang2017}.
When $\delta=\frac\pi6$, they become the three minima of the UUD phase, and the state of UUD phase will spontaneously break this discrete $C_3$ symmetry.

\begin{figure}[htb]
\includegraphics[width=0.95\columnwidth]{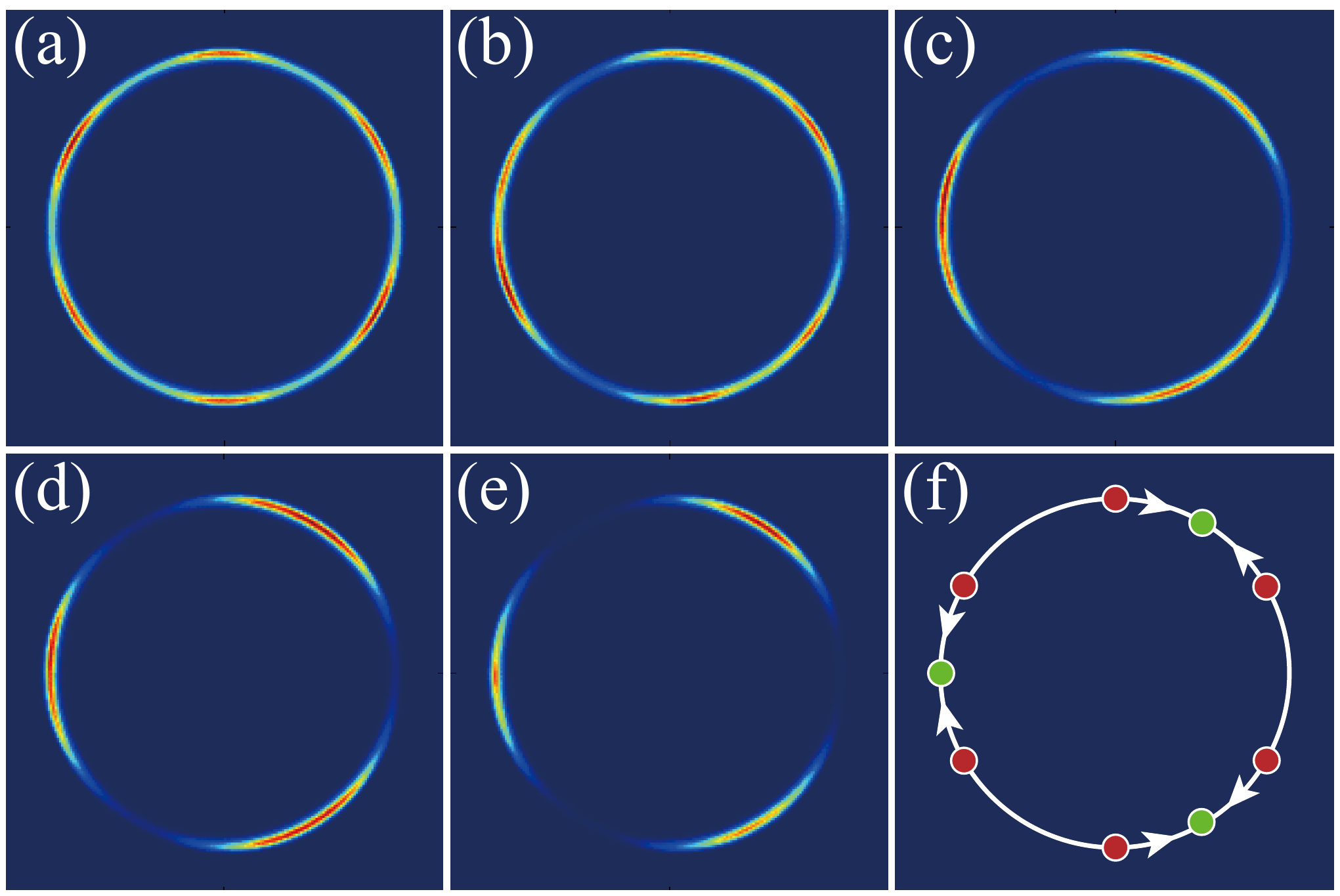}
\caption{(a)-(e) Histogram of $me^{i\theta}$ from $h=0$ to $h=0.01$ with interval $\Delta h=0.0025$. In QMC simulations, the temperature is $T=1/64$ and the system size is $L = 9$, the parameter path is from the clock phase to the UUD phase indicated by the red dashed arrow in the inset of Fig.~\ref{fig:fig1}(b). (f) The six red points represent the locations of six minima of complex order parameter in clock phase. The three green points represent the locations of three minima of complex order parameter in UUD phase. Their movements (indicated by the white arrows) are shown in (a) - (e) with real QMC data.}
\label{fig:fig2}
\end{figure}

\begin{figure}[htb!]
\includegraphics[width=0.9\columnwidth]{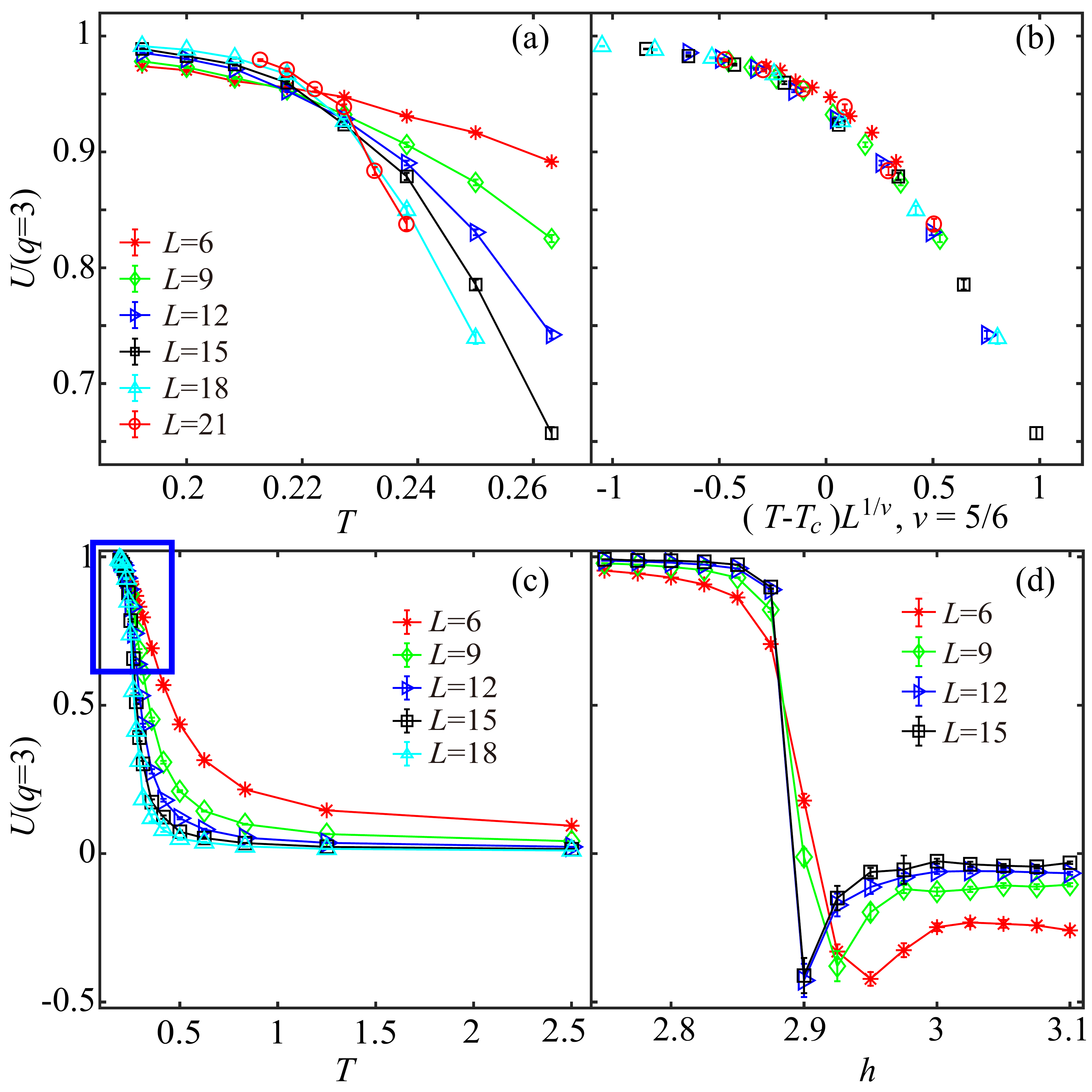}
\caption{(a) shows the Binder cumulant $U(q)$ of the 2D $q=3$ Potts transition. At $h=0.5$ and the temperature reduces from the PM to UUD phases. The cross of $U$ for different system sizes indicate the continuous phase transition point between PM at high temperature and UUD phase at low temperature, with $T_c=0.23(2)$. In (b), we perform the data collapse of Binder cumulant with 2D $q=3$ Potts critical exponent $\nu=5/6$ as $(T-T_c)L^{1/\nu}$ close to $T_c$. (c) shows the wider temperature range of the Binder cumulant compared with the zoom-in in (a) -- denoted as the blue rectangle box. $U$ varies continuously from 0 at high temperature to 1 at low temperature. The nature of this continuous transition is revealed in (a) and (b). In (d), along the path of black dashed arrow in Fig.~\ref{fig:fig1}(b), the transition from the UUD phase to the polarized PM phase is weakly first-order, as indicated by the negative dip in the $U$.
}
\label{fig:fig3}
\end{figure}

It is interesting to see that such symmetry analysis can be verified by QMC simulation.
As shown in Fig.~\ref{fig:fig2}(a)-(e), we plot the histograms of the complex order parameter $me^{i\theta}$
at different $h$ [starting from $h=0$ and increasing with footstep $\Delta h=0.0025$ to $h=0.01$,
the path is shown by the red dashed arrow in the inset of Fig.~\ref{fig:fig1}(b)] at low temperature
$T=1/64$ with system size $L=9$. The positions of bright spots represent
the locations of minima, and they move as expected.
Fig.~\ref{fig:fig2}(f) shows the locations of minima of $me^{i\theta}$ in theoretical symmetry
analysis and the white arrows indicate the movement of the bright spots. The finite (but low) temperature continuous transition between the clock and
the UUD phases therefore belongs to the 2D Ising universal class,
as each green dot minimum of $\theta$ further splits
into two red ones. On the other hand, the continuous transition from the PM to the UUD phases,
as we will discuss below, belongs to the 2D $q=3$ Potts universal class,
as $\theta$ falls into one of the three green dot minima.
Physically, the transitions at these two phase boundaries break the three-fold rotation
and mirror reflection symmetries of the lattice, respectively.

\section{Phase diagram}
Our phase diagram is shown in Fig.~\ref{fig:fig1}(b), the PM, UUD, BKT and clock phases are in place.
Our numerical results reveal that the UUD-PM phase transition is continuous and belongs to the 2D
$q=3$ Potts universality at weak field ($h<1.5$) and becomes weakly first order at strong field $(h>1.5)$~\cite{takahiro2015,yuzhi2020} (see Appendices~\ref{app:tricritical} for details). The boundary of UUD phase
extrapolates to the upper BKT temperature at $h=0$.

To study the UUD-PM phase transition and its finite size scaling in QMC,
we employ the Binder cumulant for the $q$-state Potts phase transition,
$U(q) = \frac{q+1}{2}(1-\frac{q-1}{q+1}\frac{\langle m^4 \rangle }{\langle m^2 \rangle^2})$,
where $m$ is the order parameter,  $\langle \ \rangle$ means the statistical average in QMC.
For a continuous transition, $U$ extrapolates to the saturation value $1$ when $T<T_c$,
and decays to $0$ at high temperature when $T>T_c$, and curves of different sizes cross at $T_c$~\cite{Binder-1981a, Binder-1981b}.
This is what we saw in the Fig.~\ref{fig:fig3}(a) and (c), in which the QMC results of $U(q=3)$
along the path of blue dashed arrow in Fig.~\ref{fig:fig1}(b) are shown.
Fig.~\ref{fig:fig3}(c) is the overall scale with a wide temperature range
and we note that $U(q=3)$ indeed reduce from $1$ to $0$ as temperature increases.
Fig.~\ref{fig:fig3}(a) is a zoom-in in the critical region (as denoted by the blue rectangle in Fig.~\ref{fig:fig3}(c)),
here $U$ crosses at a critical point for different $L$ and one can readily read $T_c = 0.23(2)$.
We further collapse $U$ with the 2D $q=3$ Potts exponent $\nu = 5/6$ according to the
scaling relation $U(q) = f((T-T_c)L^{1/ \nu})$, and the results are shown in Fig.~\ref{fig:fig3}(b).
This good data collapse evidence that UUD-PM transition is continuous
and indeed belongs to the $q=3$ Potts universality. The other blue square phase boundary
points in Fig.~\ref{fig:fig1}(b) are determined in similar manner.

\begin{figure}[htb]
\includegraphics[width=\columnwidth]{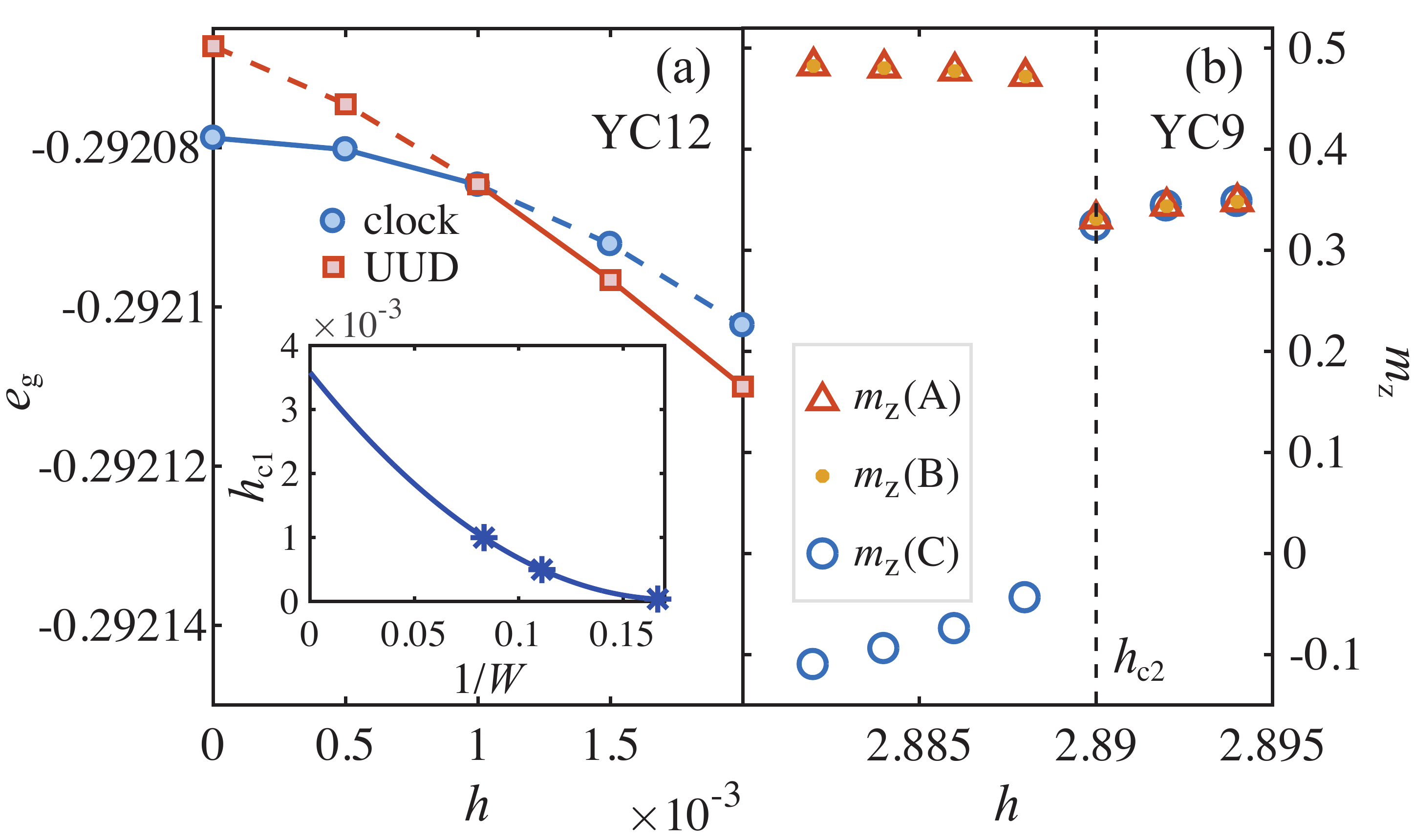}
\caption{(a) The bulk energy curves $e_g$ calculated by
DMRG on the YC$12\times30$ cylinder under
different (clock and UUD) boundary pinning fields
exhibit a crossing at $h_{c1} \simeq 0.001$.
In the inset, the crossing fields $h_{c1}$
calculated under system width $W=6$, 9 and 12
are extrapolated to $1/W\to 0$, from which we
determine $h_{c1}\simeq 0.0036$
as the transition point between the two phases.
(b) The QPT near the upper critical field
$h_{c2} \simeq 2.89$. The magnetization $m_z$
results provide clearly evidences of a three-sublattice
UUD order for $h < h_{c2}$, while $m_z$ becomes uniform
in the polarized phase for $h_ > h_{c2}$. The data
are from DMRG calculations on a YC$9\times45$ cylinder.}
\label{fig:DMRG}
\end{figure}

At large $h$, the transition between UUD and PM becomes weakly first order,
 and this can also be seen from the Binder cumulant~\cite{Binder1984}.
As shown in Fig.~\ref{fig:fig3}(c) and (d), we make a comparison between
the Potts transition and the first order transition, along the two different
parameter paths indicated by the blue and black dashed arrows in Fig.~\ref{fig:fig1}(b).
In Fig.~\ref{fig:fig3}(c) the continuous 2D Potts transition is shown,
and one sees that the $U$ changes continuously and always above zero, and
the different curves cross at the $T_c=0.23(2)$ (as shown in Fig.~\ref{fig:fig3}(a)).
The contrast is very obvious in Fig.~\ref{fig:fig3}(d). Here we fix the temperature
at $T=1/32$ and scan the field $h$. It is clear that close to the UUD to PM
(here means polarized phase) transition, the $U$ develops a negative dip
at the transition point, this is the signature of a first-order phase transition,
where the system in the Monte Carlo simulations can be either inside the ordered phase or inside the disorderd phase~\cite{Binder1984}.
And when the $h>2.9$, the system enters the PM phase.
We note that the transition is acutually of weakly first order,
as the $U$ didn't diverge at the present $L$, this is also consistent
with the experiment in TmMgGaO$_4$ with the magnetic excitation
gap observed to be very small near the high transition field~\cite{Qin2020fieldtuned}.

With the high field phase boundary between
the UUD and PM determined, we now look for the low field
ones between the clock phase and the UUD phase, as indicated in the inset of Fig.~\ref{fig:fig1}(b).
The RG calculation in Appendices~\ref{app:RG} predicted the shape of the phase boundary boundary
and find it lies very close to the vertical $T$ axis with $h=0$ (see Fig.~S3),
and our numerical result is consistent with such prediction. The symmetry analysis reveals
that this is an Ising transition since the clock and UUD phases differs by a $Z_2$ symmetry
(cf. Fig.~\ref{fig:fig2}). Since the temperature here turns out to be too low to have
large-scale QMC simulations, we turn to DMRG computation directly at $T=0$.

\section{DMRG study of the ground-state phase diagram}
DMRG calculations are employed to compute the groundstate
phase diagram.
In Fig.~\ref{fig:DMRG}, we show the numerical evidence of two
QPTs at lower ($h_{c1}$) and higher ($h_{c2}$) critical fields, respectively.
In the calculations, we introduce two kinds of pinning fields on the
outmost two columns of the cylindrical system, which respectively
favor the clock and UUD order, and compute the observables in
the bulk (see Appendices~\ref{app:DMRG}).
Such pinning fields, although applied only on the boundary
of the system, can nevertheless stabilizes the two types of ordered
states in the bulk, and can be used to discern the subtle
competition between different quantum states at low magnetic fields.

In Fig.~\ref{fig:DMRG}(a), we plot the bulk energy measured
in the very center of the long cylinder, and see that the clock phase
is energetically more favorable at lower fields, while the UUD phase
takes the place as the true ground state at higher fields,
with the crossing point at $h_{c1}\simeq 0.001$ (for width $W=12$),
which could be extrapolated to $0.0036$ for $1/W \to 0$,
as shown in the inset of Fig.~\ref{fig:DMRG}(a). As the field further increases,
the UUD order would also get suppressed, and a phase transition to polarized phase
takes place at around $h_{c2}\simeq 2.89$. As shown in Fig.~\ref{fig:DMRG}(b),
the local magnetization in the UUD phase, $m_z(A)=m_z(B) \neq m_z(C)$
on the three sublattices A, B, and C, becomes uniform $m_z(A)=m_z(B)= m_z(C)$
in the polarized phase. The obtained ground state $h_{c1}$ and $h_{c2}$
are denoted as the yellow stars in the phase diagram of Fig.~\ref{fig:fig1}(b),
and are well connected with the finite temperature phase boundaries of QMC.

\section{Experimental relevance}
Our results have immediate connections to experiments on the
rare-earth compound TmMgGaO$_4$~\cite{Cava2018,Shen2019,Li2020,Lih2020,ZHu2020},
where the existence of BKT transitions and and an intermediate floating phase has
been theoretically proposed~\cite{Lih2020} and experimentally
detected in NMR~\cite{ZHu2020}.
There is currently upsurging research interest on further experimental exploration
of the BKT physics~\cite{Dun2020neutron}, and the field-tunable quantum states
in the compound~\cite{Qin2020fieldtuned}. Our results, combining finite-temperature
QMC and ground-state DMRG calculations, unambiguously pinpoint the two quantum
phase transitions, at the lower- $h_{c1}$ (clock to UUD) and upper-field $h_{c2}$
(UUD to polarized). Although the model parameters in this work are simplified~\cite{Lih2020},
we can still resolve the two QPTs and the upper field $h_{c2}$ is very close
to 3.7~T when taking $J_1=1$~meV, in good agreement to
experiments~\cite{Shen2019,Li2020,ZHu2020}.

Moreover, we show numerical evidence from large-scale QMC calculations that there exist
a continuous thermal phase transition that belongs to the 3-state Potts universality class,
under finite external fields. Such Potts transition gradually changes into a weakly first-order
phase transition around $h_{c2}$. These theoretical predictions can be confirmed in future
thermodynamic and dynamical measurements of the material, which should exhibit universal scaling near the
transition point. Our model could also be implemented and further detected in the
programmable quantum simulators based on Rydberg atom arrays ~\cite{Rhine2021,Ebadi2020,scholl2020programmable}
and superconducting qubits~\cite{King2018topology,King2019scalingadv} where geometry
frustration and quantum dynamics of quantum Ising models have been proposed
and partially realized.

\section*{Acknowledgments}
We thank Weiqiang Yu, Jinsheng Wen for valuable discussions. YDL, ZY and ZYM acknowledge support from the RGC of Hong Kong SAR of China (Grant Nos. 17303019 and 17301420), MOST through the National Key Research and Development Program (Grant No. 2016YFA0300502) and the Strategic Priority Research Program of the Chinese Academy of Sciences (Grant No. XDB33000000). HL and WL acknowledge the support from the National Natural Science Foundation of China (Grant Nos. 11834014 and 11974036). HTW and QY acknowledge supports from MOST under Grant No. 2015CB921700 and from NSFC under Grant No. 11874115.
We thank the Center for Quantum Simulation Sciences in the Institute of Physics, Chinese Academy of Sciences, the Computational Initiative at the Faculty of Science and the Information Technology Services at the University of Hong Kong and the Tianhe-1A Tianhe-2 and Tianhe3 prototype platforms at the National Supercomputer Centers in Tianjin and Guangzhou for their technical support and generous allocation of CPU time.


\appendix

\section{SSE-QMC Method}
\label{app:SSE}

In this section, we describe the implementation of SSE-QMC algorithm of the quantum Ising model~\cite{sandvikTFIM,melko2013stochastic,JRZhao2020}.
\subsection{SSE on $\sigma^{z}$ Basis}
The Hamiltonian for our model is $H = J_1\sum_{\langle i,j \rangle} S_i^z S_j^z + J_2 \sum_{\langle\langle i,j \rangle\rangle} S_i^z S_j^z \nonumber - \sum_i (\Delta S_i^x + h S_i^z)$
, we can decompose this Hamiltonian into diagonal or off-diagonal site and bond operators
\begin{equation}
\begin{split}
H_{0, 0}&=I\\
H_{-1,i}&=\frac\Delta 2(S_{i}^{+}+S_{i}^{-})\\
H_{0,i}&=h_0 + h S_{i}^{z}\\
H_{1,b}&=J_1 (\frac{1}{4}-S_{b(1)}^{z} S_{b(2)}^{z})\\
H_{1,b'}&=J_2 (\frac{1}{4}-S_{b'(1)}^{z} S_{b'(2)}^{z})
\end{split}
\end{equation}
with $H =-\sum_{i}(H_{-1,i}+H_{0,i}) - \sum_{b,b'}(H_{1,b}+H_{1,b'})$, noting we add some constants into it which doesn't change the physics.
Here $b$ ($b'$) means the NN (NNN) bond, $H_{0,0}$ denotes the identity operator, $H_{-1,i}$ ($H_{0,i}$) denotes the off-diagonal (diagonal) operator on site $i$, $H_{1,b}$ ($H_{1,b'}$) denotes the diagonal operator on bond $b$ ($b'$). $b(1)$ ($b'(1)$) and $b(2)$ ($b'(2)$) represent the two sites connecting the bond $b$ ($b'$).
$S_{i}^{+}=S_{i}^{x}+iS_{i}^{y}$ and $S_{i}^{-}=S_{i}^{x}-iS_{i}^{y}$ represent the creation and destruction operator of spin-1/2.
In order to make these site and bond operators positive, we add $\frac{1}{4}$ in $H_{1,b}$ ($H_{1,b'})$ and have to set the constant $h_0 \geq h/2$.

It is well known that the partition function $Z=\mathrm{Tr}\,e^{-\beta H}$ can be expressed as a power series expansion:
\begin{equation}
Z = \sum\limits_\alpha \sum_{S_M} {\beta^n(M-n)! \over M!}
    \left \langle \alpha  \left | \prod_{i=1}^M H_{a_i,p_i}
    \right | \alpha \right \rangle ,
\label{zm}
\end{equation}
where $\beta$ is the inverse of temperature, and $M$ is the truncation of the expansion series $n$. Taking $S^{z}$ as a complete set of basis for the system, the non-zero matrix elements for site operators and bond operators are
\begin{equation}
\begin{split}
\langle \uparrow|H_{-1,i}|\downarrow\rangle &= \langle \downarrow|H_{-1,i}|\uparrow\rangle=\frac\Delta 2 \\
\langle \uparrow|H_{0,i}|\uparrow\rangle &= h_0+\frac h 2\\
\langle \downarrow|H_{0,i}|\downarrow\rangle &= h_0-\frac h 2\\
\langle \downarrow\uparrow|H_{1,b}|\downarrow\uparrow\rangle &=\langle \uparrow\downarrow|H_{1,b}|\uparrow\downarrow\rangle=\frac {J_1} 2 \\
\langle \downarrow\uparrow|H_{1,b'}|\downarrow\uparrow\rangle &=\langle \uparrow\downarrow|H_{1,b'}|\uparrow\downarrow\rangle=\frac {J_2} 2 \\
\end{split}
\end{equation}

\begin{figure}[htp]
	\centering
	\includegraphics[width=0.5\columnwidth]{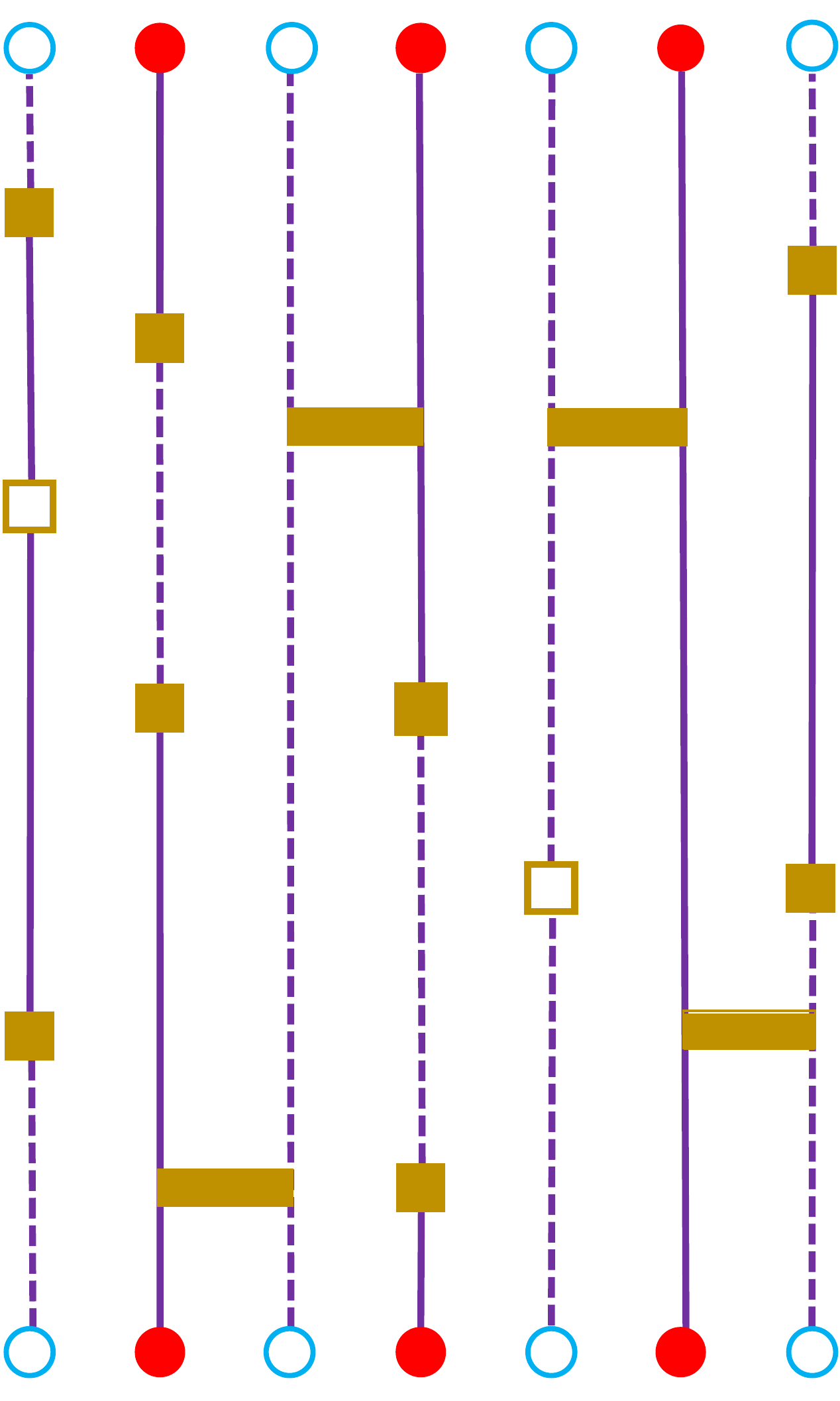}
	\caption{SSE-QMC configuration of quantum Ising model. Golden bars represent Ising bond operators. Filled squares plaquette are off-diagonal site operators, and open plaquettes denote the diagonal site operators. Arrows represent periodic boundary conditions in the imaginary time direction. The red solid circles and the light blue open circles indicate spin up and down. Solid and dashed purple lines illustrate the spin states (spin up or down).}
	\label{fig:fig7}
\end{figure}

The updating scheme includes the diagonal update which either inserts or removes a diagonal operator between two states with probabilities regulated by the detailed balanced condition, and the cluster update which flips all the spins as following scheme. The configurations of the updating scheme are shown in Fig.~\ref{fig:fig7}.

We describe the updating scheme in the following steps:
\begin{enumerate}
	\item \textbf{Diagonal update}\\
	We go through the operator strings and either remove or insert a diagonal operator according to the following procedures.
	\begin{enumerate}
		\item For a null operator ($H_{0,0}$) , we substitute it with a diagonal operator $H_{0,i}$ $H_{1,b}$ or $H_{0,b'}$  by the procedures below.
			Firstly we make the decision of which position for diagonal operators to insert with a probability $1/(N+N_b+N_{b'})$, where $N$ means the number of total sites, $N_b$ ($N_{b'}$) means the number of total NN (NNN) bonds. If the chosen bond to insert a bond operator has an anti-parallel configurations, then the insertion  of a bond operator at this place is prohibited. After the decision is made, accept the insertion of an operator with probability
			\begin{equation}
			P=\min\left(\frac{\beta (N+N_b+N_{b'}) \langle H_a \rangle}{M-n},1\right),
			\end{equation}
			The $H_a$ means $H_{0,i}$ $H_{1,b}$ or $H_{0,b'}$ which depends on the position we chosen.
		\item For a diagonal operator $H_a$,
		we removed  it with probability, \begin{equation}
		P=\min\left(\frac{M-n+1}{\beta(N+N_{b}+N_{b'})\langle H_a \rangle},1\right)
		\end{equation}

	\item For an off-diagonal operator, we ignore it and go to the next operator in the operator strings.
	\end{enumerate}
	\item \textbf{Cluster update}
	\begin{enumerate}
		\item We generally follow two rules to construct the clusters: (1) clusters are terminated
		on site-operators $H_{-1,i}$ or	$H_{0,i}$; and (2) the four legs of a bond operator
		$H_{1,b}$ ($H_{1,b'}$) belong to one cluster.
		Carry out this procedure until all the clusters are bulit, and a configuration of clusters are shown in Fig.~\ref{fig:fig7}.
		\item Due to the external field breaks $Z_2$ symmetry, the probability of cluster update should be modified~\cite{OFS2002,yan2019sweeping,yan2020improved}. Clusters identified from the above rules are then flipped with probability $\frac{W_{new}}{W_{old}+W_{new}}$. Here the $W_{new/old}$ is the weight after/before cluster update.
	\end{enumerate}

\end{enumerate}

\section{$q=3$ Potts to first order transition}
\label{app:tricritical}

\begin{figure}[htb]
\includegraphics[width=0.95\columnwidth]{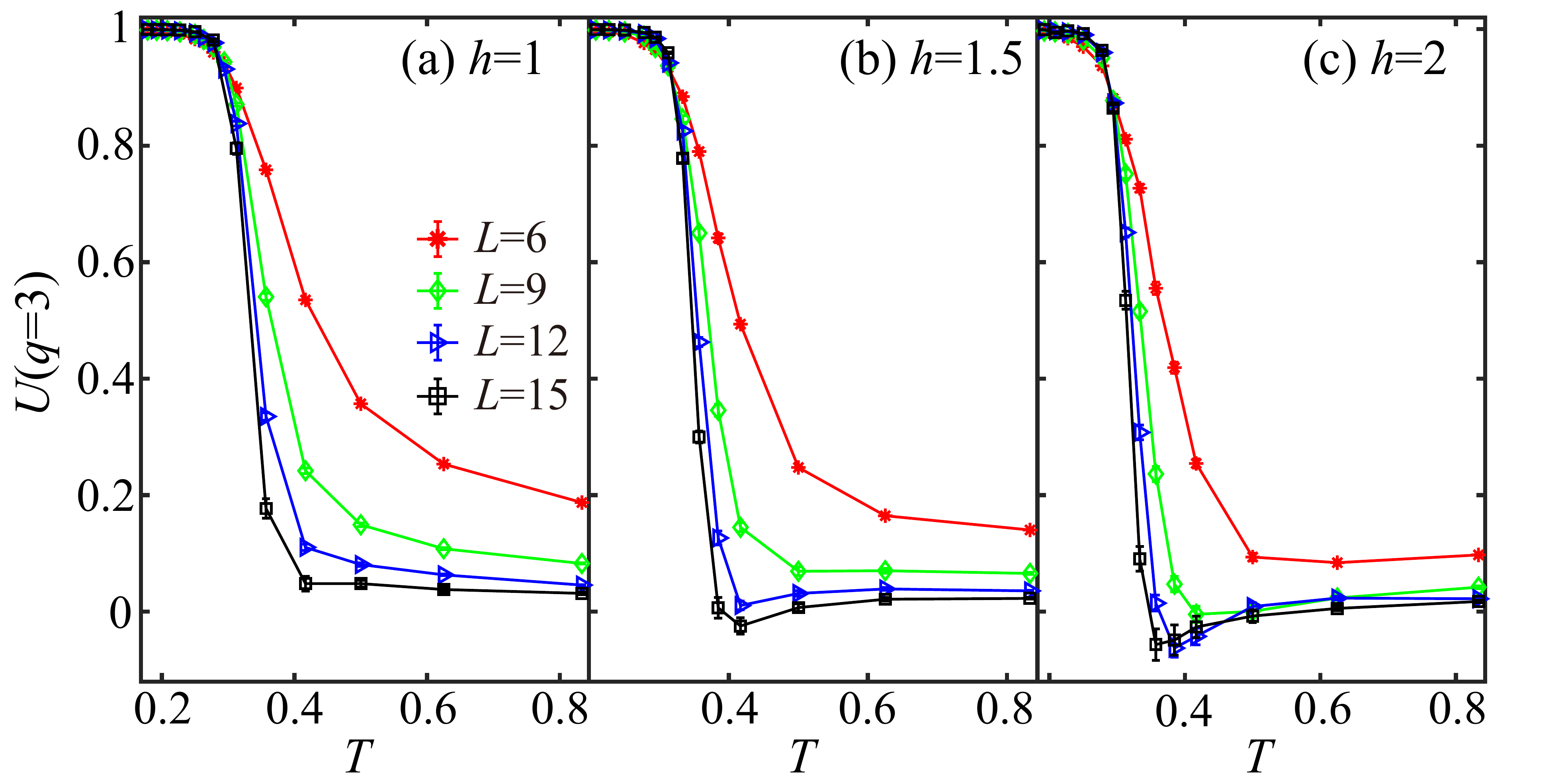}
\caption{Binder cumulant $U(q=3)$ of the $q=3$ Potts transition at (a) $h=1$, (b) $h=1.5$ and (c) $h=2$.}
\label{fig:tricritical}
\end{figure}

As mentioned in the main text, the UUD-PM phase transition is continuous when the external magnetic is relatively small ($h<1.5$), and the UUD-PM phase transition becomes the first order when the external magnetic is large enough ($h>1.5$). To prove that, we plot the 3-states Binder cumulant $U(q=3)$ as function of temperature $T$ at different external magnetic $h=1$, $h=1.5$ and $h=2$ for different system sizes $L=6,9,12,15$, as shown in Fig.~\ref{fig:tricritical}. We notice that the $U(q=3)$ decay smoothly at $h=1$ for $L=6,9,12,15$, however $U(q=3)$ begins to develop a negative dip at $h = 1.5$ for $L=12$ and $L=15$, which is the typical behavior close to a first-order phase transition~\cite{Binder1984,yuzhi2020,takahiro2015}. When $h=2$, the dips go to more negative, which means the first-order phase transition is more noticeble.

However, we note that the negative dip in the Binder cumulant along our phase boundary has not developed into the divergence which is usually the case for the standard first order phase transition, this is consistent with the fact that the 3D $q=3$ Potts (or its 2D quantum version in the present case at $T=0$ from UUD to PM) is known to be a weakly first order, in fact, "almost second order" phase transition~\cite{Bloete1979}.

\section{RG Analysis}
\label{app:RG}

\graphicspath{{./Figs/}}

\newcommand{\fig}[1]
{Fig.~\ref{#1}}
\newcommand{\eq}[1]
{Eq.~\ref{#1}}
\newcommand{\kb}[0]
{k_{\rm B}}

In this section, we describe how to construct the Ginzburg-Landau theory of
the 3-sublattice order parameter to the triangular lattice quantum Ising model, and analyze the phase transitions at very small external field by renormalization group (RG) methods.

\subsection{Ginzburg-Landau Theory}

To study the low-energy effective theory of the triangular lattice transverse field Ising model, we start from the formulation of the three-sublattice ordering standard Ginzburg-Landau (GL) theory. To capture physics of the three-sublattice ordered structure, the order parameter is defined at the $\bm{K}$ point:
\begin{equation}\label{Eq:orderparameter}
    \psi(\bm{r}) = |\psi(\bm{r})| e^{i\theta_{\bm{r}}} = \sum_i m^z_i e^{i\bm{K}\cdot\bm{R}_i},
\end{equation}
where $\bm{K} =(2\pi/3,2\pi/3)$ is the corner of the hexagonal Brilliouin zone (BZ), and $\bm{R}_i, i=1,2,3$ represents coordinates at three sublattice sites in the unit cell at $\bm{r}$.

Traditional symmetry analysis \cite{Domany1978,Domany1979,Alexander1975,Damle2015,Biswas2018} provides a coarse-grained oder parameter field Ginzburg-Landau free energy density\cite{Biswas2018}:

\begin{align}\label{eq:GLtheory}
    \mathcal{F}_{ts}[\psi]= & \kappa |\nabla\psi|^2+r|\psi|^2+u_4|\psi|^4+u_6|\psi|^6 \nonumber              \\
                            & +\lambda_6|\psi|^6 \cos(6\theta)+\lambda_{12}|\psi|^{12}\cos(12\theta)+\cdots.
\end{align}

In the phase transitions we want to study, we expect local amplitude of the order parameter $|\psi|$ retains nonzero and slowly-variant, and the fluctuation of phase $\theta_{\bm{r}}$ is dominant. From the following analysis, we will see why $\lambda_{12}$ term is not important. These result in the further simplification of the model Eq.~\eqref{eq:GLtheory} into
\begin{equation}\label{eq:GLZeroField}
    \mathcal{F}_{ts}[\theta_{\bm{r}}]=\frac{1}{4\pi g}(\nabla\theta)^2+\lambda\cos(6\theta),
\end{equation}
where $g(T)$ is monotonously tuned by temperature $T$, and $\lambda$ is the rescaled coupling constant of $\cos(6\theta)$ term.

From the definition of the three-sublattice order parameter, i.e. in Eq.~\eqref{Eq:orderparameter}, we find when $\lambda>0$ the phase $\theta$ falls in 6 minima that represent a six-fold degenerate clock phase, while $\lambda<0$ represents the up-up-down (UUD) phase with nonzero net magnetization. We expect similar critical behaviors to 6-state clock model \cite{Jose1977,Challa1986}.

A further observation \cite{Damle2015}, however, pointed out that the above three sublattice order parameter doesn't capture the physics of uniform magnetization mode, like in UUD phase, at $\Gamma$ point:
\begin{equation}
    m_{\bm{r}}\equiv\sum_i m^z_i.
\end{equation}
Regarding the same $S_3\times Z_2$ symmetry as the original GL theory when the net magnetization is present, a coarse-grained Ising field, and - most importantly - a coupling between three sublattice phase mode and uniform magnetization Ising mode, are included in the model:
\begin{equation}\label{eq:GLw/Ising}
    \mathcal{F}[\theta_{\bm{r}},m_{\bm{r}}]=\mathcal{F}_{\mathrm{ts}}[\theta_{\bm{r}}]+\mathcal{F}_{\mathrm{Ising}}[m_{\bm{r}}]-cm_{\bm{r}}\cos(3\theta_{\bm{r}}).
\end{equation}

The last term, when uniform magnetization is linearly polarized by the external uniform longitudinal field $h$ and thus the $Z_2$ symmetry is broken, becomes a coupling between $h$ and $\cos(3\theta)$. Focusing on the phase degree of freedom, the GL theory is written as:
\begin{equation}\label{eq:GLw/field}
    \mathcal{F}[\theta_{\bm{r}},h]=\frac{1}{4\pi g}(\nabla\theta)^2+\lambda\cos(6\theta)-h\cos(3\theta),
\end{equation}
here we rescale $h$ to absorb nonuniversal coupling coefficient, and its sign (the direction of external field) selects which 3 minima are the ground states.

\subsection{RG Equations}

The model Eq.~\eqref{eq:GLZeroField} proposed above is well-known for the existence of a power-law correlation quasi long-range ordered intermediate Berezinskii-Kosterlitz-Thouless (BKT) phase \cite{Sachdev}:
\begin{equation}\label{eq:PwrCorr}
    \left< e^{i(p\theta(\bm{r})-p'\theta(0))}\right>\sim r^{-p^2g}\delta_{pp'},
\end{equation}
when $g(T)=\eta(T)\in(\frac{1}{9},\frac{1}{4})$ for the triangular lattice quantum Ising model~\cite{Isakov2003}.

In the renormalization group (RG) language, the gradient term in Eq.~\eqref{eq:GLZeroField} is exactly marginal along the entire $\lambda=0$ line \cite{Fradkin,NagaosaSCES,FrancescoCFT,HenkelCFT}. The six-fold anisotropy term is relevant only when $g<\frac{1}{9}$, and symmetry-breaking long-range order is thus formed at low $T$; when $g>\frac{1}{4}$ the vortex term comes in, and the vortex proliferate phase arises at high $T$ \cite{Jose1977}. The intermediate BKT phase is a critical phase formed by a line of central charge $c=1$ fixed points \cite{Damle2015}, where no anisotropy is relevant and an emergent $U(1)$ symmetry is present.

The above 1st order RG analysis is performed in zero field. When nonzero uniform magnetic field is present, however, the $h\cos(3\theta)$ term is relevant everywhere $g<\frac{2}{9}$, from zero temperature to somewhere deep in disordered phase. To study how this term interacts with six-fold anisotropy term and vortex term, a detail 2nd order analysis is needed.

From the momentum shell RG, if we nondimensionalize coupling constants of every term in the theory and write it as
\begin{equation}
    S=\int d^Dx\sum_n \nu_n \Lambda^{D-\Delta_n} \phi_n(\bm{x}),
\end{equation}
where $D$ is the spatial dimension, $\Lambda$ is the large-momentum cutoff, and $\Delta_n$ and $\nu_n$ is the scaling dimension and dimensionless coupling constant of operator $\phi_n$, the 2nd order general RG equation involving operator product expansion (OPE) is \cite{Fradkin}
\begin{equation}\label{eq:genRG}
    \frac{d\nu_k}{dl}=(D-\Delta_k)\nu_k-\frac{S_D}{2}\sum_{nm}\nu_n\nu_mC_{nmk}+\cdots=\beta(\nu_k).
\end{equation}
Here $C_{nmk}$ is the structure constant of OPE, and $S_D$ is the spherical phase factor in $D$-dimension.

We first study the RG equations near the lower BKT critical point $T_L$, where $g_L\equiv g(T_L) = \frac{1}{9}$ and six-fold anisotropy term is marginal. In the $D=2$ Kosterlitz RG we are interested here, we denote $\phi_0$ the operator $(\nabla\theta)^2$, and $\phi_p$ the vortex operator $e^{ip\theta}$. The scaling dimensions $\Delta_0=2$, and $\Delta_p=\frac{p^2g}{2}$ from Eq.~\eqref{eq:PwrCorr}, resulting in the aforementioned 1st order analysis of six-fold degenerate order and relevance of external field coupling \footnote{$h\rightarrow h \Lambda^{3/2}$ makes $h$ a dimensionless quantity so that it is comparable with unity, and all other terms are already marginal.}.

The OPE calculated by 2D free boson CFT \cite{Fradkin,FrancescoCFT} tells us \footnote{The second structure constant is valid only when $n,m\neq0$, for only vortex operator calculation involved.}:
\begin{align}\label{eq:StructConst}
    C_{n,-n,0} & =-\pi\Delta_n, & C_{n,m,n+m} & =1.
\end{align}
When only one pair of vortex operators $\phi_{\pm n}$ are in the place, only the first structure constant is nontrivial to the 2nd order. The substitution of Eq.~\eqref{eq:StructConst} into the general RG equation Eq.~\eqref{eq:genRG} results in the well-known hyperbolic RG flows of sine-Gordon model. However, when there are more than one pair of vortex operators --- exactly the case for our nonzero field theory --- the coupling between constants plays an important role in determining the nature of the phase diagram.

Investigate Eq.~\eqref{eq:GLw/field}, with straightforward substitutions of nonzero coupling constants into Eq.~\eqref{eq:genRG}, and we write out RG equations when both $\cos(6\theta)$ and $\cos(3\theta)$ term are present in the theory:

\begin{align}
    \frac{d\lambda}{dl} & =(2-18g)\lambda-\frac{\pi}{2} h^2,  \label{eq:RGlambda}                             \\
    \frac{dh}{dl}       & =(2-\frac{9}{2}g - \pi \lambda )h, \label{eq:RGh}                                   \\
    \frac{dg}{dl}       & =-9\pi^2 g^3(4\lambda^2+h^2)\simeq-\frac{\pi^2}{81}(4\lambda^2+h^2). \label{eq:RGg}
\end{align}

The above equations shows the existence of a Ising class phase transition. While the strong relevance of $h\cos(3\theta)$ term drives the parameter flow to $\lambda\rightarrow-\infty$, ie. UUD phase, in most part of the parameter space, a parameter flow starting from $\lambda\gg |h|$ when $g<g_L$ will suppress the increase of $h$ according to Eq.~\eqref{eq:RGh}, thus in turn go to $\lambda\rightarrow\infty$ the clock phase. Therefore, if the system is in clock phase when temperature is low enough and no external field applied, it will undergo an Ising phase transition into UUD phase at some finite field strength when the external uniform magnetic field is turned on and increased. And $g_L$, $h=0$ is a tricritical point, where clock phase, UUD phase and BKT phase boundaries merges at one point.

Here we complete the RG analysis near $g_L$ and we now turn to study the phase transition near upper BKT temperature $g_U\equiv g(T_U) = \frac{1}{4}$. In this regime the six-fold anisotropy is no longer relevant, and the relevance of the vortex term $y\cos(\phi)$ defined on the dual lattice interacts with the external Field term. Given the complexity of duality mapping, the result is expected to be complicated. However, \cite{Jose1977} gave a concise argument that up to the 2nd order, the coupling constants $h$ and $y$ don't interact:

\begin{align}
    \frac{dh}{dl} & =(2-\frac{9}{2}g)h, \label{eq:RGhUpper}                \\
    \frac{dy}{dl} & =(2-\frac{1}{2g})y, \label{eq:RGy}                     \\
    \frac{dg}{dl} & =\frac{\pi^3}{g}y^2-9\pi^2 g^3h^2. \label{eq:RGgUpper}
\end{align}
It is based on two facts: the relation when doing duality mapping between lattices:
\begin{align}
    9g & \Leftrightarrow\frac{1}{g}, & y & \Leftrightarrow h, & h & \Leftarrow y,
\end{align}
and the invariance of RG equations when $h\rightarrow-h$. If any $gh$ or $h^2$ term appears in one of Eqs.~\eqref{eq:RGhUpper}-\eqref{eq:RGy}, then a counterpart will also appear in the other equation, and it will break the invariance of RG equations under $h\rightarrow-h$.

\begin{figure}
\centering
\includegraphics[width=0.7\columnwidth]{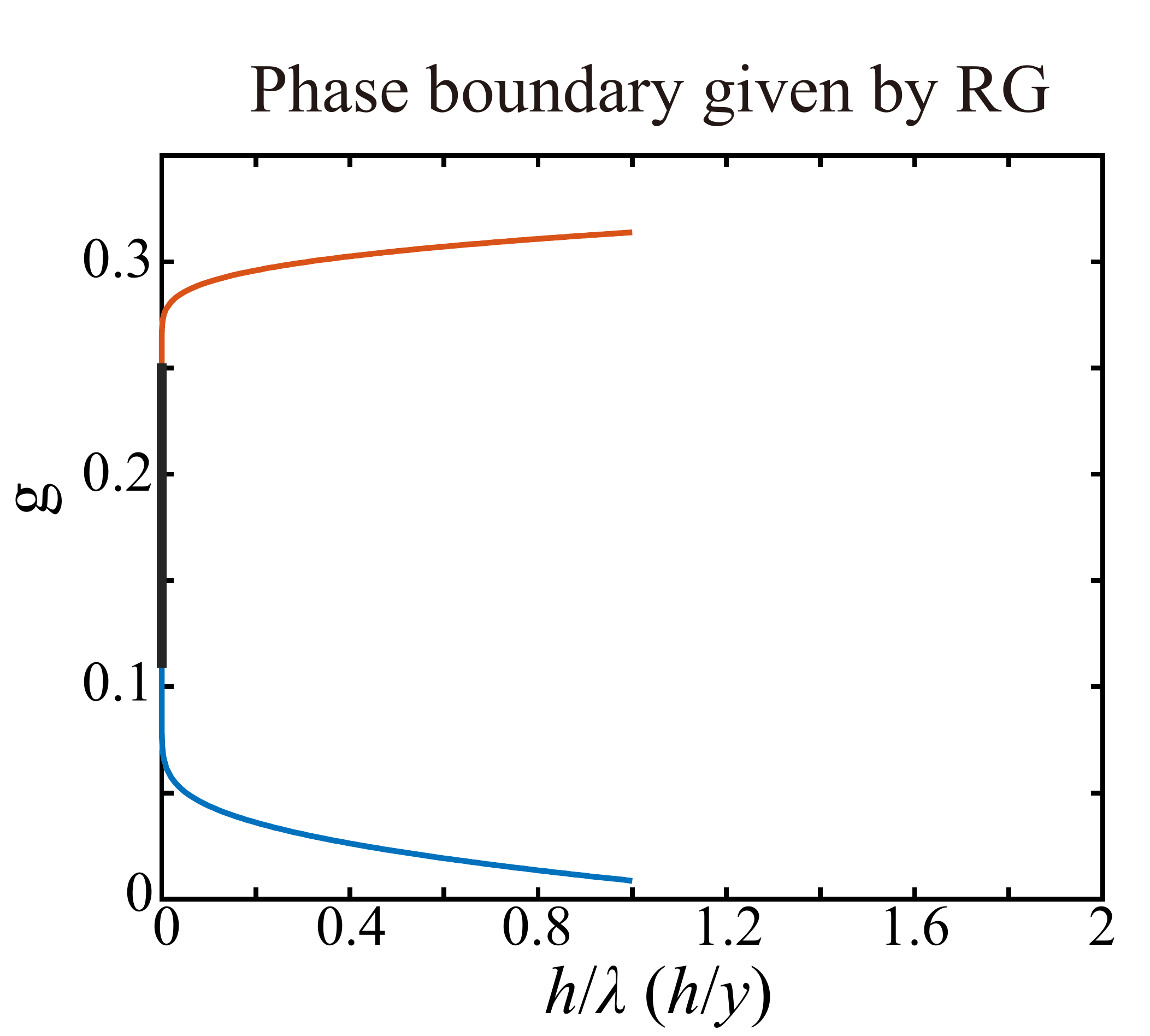}
\caption{\label{fig:PhsBdr}The phase boundaries given from RG analysis. 
Here we set dimensionless $\lambda=y=0.001$. The top orange line is where 
Potts phase transition occurs, and the bottom blue line is where Ising phase 
transition occurs.}
\end{figure}

Because $h\cos(3\theta)$ term is always relevant near the critical point, and the gapped disordered phase by the vortex term should be more robust to the perturbation than the gapless BKT phase, the phase boundary between UUD and disordered phase should be extrapolated directly to $g_U$. Therefore, the decoupled nature of multiple relevant perturbations also suggests a tricritical point at $g_U$ and $h=0$ \cite{Sachdev}, where BKT phase, UUD phase and disordered phase boundaries come into one point.

\subsection{Phase Diagram}
To see the explicit phase transition analysis of RG at the small field regime, the best way is to show it in the phase diagram.

Here we plot the obtained RG phase diagram in Fig.~\ref{fig:PhsBdr}. The parameter points in the phase diagram are the starting parameter points of RG flows. The top orange line distinguishes whether $|h|$ or $y$ goes to unity first. The former is identified as the UUD phase, and the latter the disordered phase, and the transition between them is in 3-state Potts universal class. The bottom blue line is the boundary below which $h$ no longer flows to unity but zero and is identified as clock phase, above which $|h|$ flows to unity faster than $\lambda$ and the UUD phase is clarified here. Thus this transition is in the Ising universal class.

\section{DMRG Results}
\label{app:DMRG}


\begin{figure*}[t!]
\centering
\includegraphics[width=1.6\columnwidth]{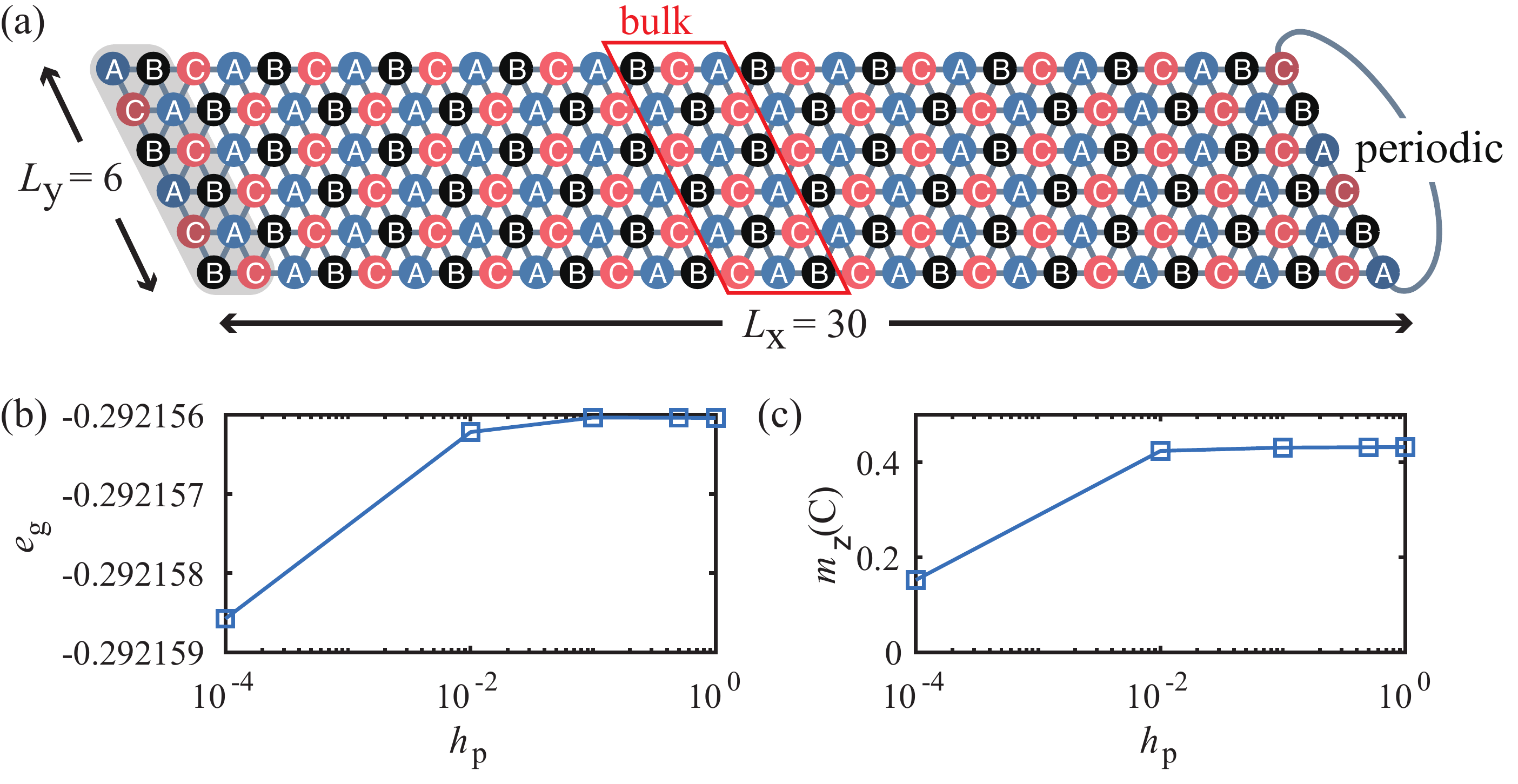}
\caption{\label{fig:DMRGSI}(a) Illustration of the three-sublattice 
spin structure on the YC6$\times$30 cylinder in the clock order phase,
with the parameters $J_1 =1$, $J_2 =0$, $\Delta=0.2$, $h=0$
and the clock-type field $h_{\rm p}=10^{-4}$ pinning on the left 
two columns of the system (the gray area). The red and blue color 
represent spin up ($m_{\rm z} > 0$) and spin down ($m_{\rm z} < 0$)
along the magnetic easy axis, and the black one stands for the 
superposition of spin up and down with zero net $z$-component 
($m_{\rm z} \simeq 0$). The brightness of the color represents the 
absolute strength of local magnetization. 
(b) shows the bulk energy $u_{\rm b}$ vs pinning field $h_{\rm p}$, 
and (c) shows the bulk magnetization $m_{\rm z}$ on the sub-lattice C. 
The bulk sites involved in the calculations are indicated in (a) within 
the red parallelogram.}
\end{figure*}

As mentioned in the main text, in the ground state
the system develops a clock order at field $h=0$,
which, however, is fragile under longitudinal fields.
To help stabilize the order in a finite-size system and 
discern the subtle competition between different magnetically 
ordered ground states, we applied a small clock-type 
pinning field $h_{\rm p}$ from very small ($10^{-4}$) 
to a moderate value ($h_{\rm p} = 1$)
on the left boundary of the system. The calculated 
on-site magnetization $m_z$ on a YC 6$\times$30 cylinder, 
as shown in Fig.~\ref{fig:DMRGSI}. Three-sublattice order 
can be clearly identified by the different colors on each site,
which represents the sign of local $m_z$. The absolute values 
of $m_z$ on sub-lattice C and A are two orders of magnitude 
larger than that on sub-lattice B, thus indicating a clock spin 
configuration. In Fig.~\ref{fig:DMRGSI}(c), we see even the 
smallest ($h_{\rm p}=10^{-4}$) field can induce a clock order, 
and the latter is converged to about 0.43  for $h_{\rm p} \geq 10^{-2}$. 
Besides, in Fig.~\ref{fig:DMRGSI}(b)  we observe that $e_g$ 
quickly converges as the pining filed $h_{\rm p}$ 
increases, and the energy result $e_g$ has converged to six digits 
for $h_{\rm p} > 0.1$. Therefore, in the main text we fix the pinning field 
$h_{\rm p}=1$ and perform the calculations at low magnetic fields 
as shown in Fig.~\ref{fig:DMRG}(a) of the main text.


%
%

\bibliographystyle{apsrev4-2}
\bibliography{main}

\begin{thebibliography}{46}%
\makeatletter
\providecommand \@ifxundefined [1]{%
 \@ifx{#1\undefined}
}%
\providecommand \@ifnum [1]{%
 \ifnum #1\expandafter \@firstoftwo
 \else \expandafter \@secondoftwo
 \fi
}%
\providecommand \@ifx [1]{%
 \ifx #1\expandafter \@firstoftwo
 \else \expandafter \@secondoftwo
 \fi
}%
\providecommand \natexlab [1]{#1}%
\providecommand \enquote  [1]{``#1''}%
\providecommand \bibnamefont  [1]{#1}%
\providecommand \bibfnamefont [1]{#1}%
\providecommand \citenamefont [1]{#1}%
\providecommand \href@noop [0]{\@secondoftwo}%
\providecommand \href [0]{\begingroup \@sanitize@url \@href}%
\providecommand \@href[1]{\@@startlink{#1}\@@href}%
\providecommand \@@href[1]{\endgroup#1\@@endlink}%
\providecommand \@sanitize@url [0]{\catcode `\\12\catcode `\$12\catcode
  `\&12\catcode `\#12\catcode `\^12\catcode `\_12\catcode `\%12\relax}%
\providecommand \@@startlink[1]{}%
\providecommand \@@endlink[0]{}%
\providecommand \url  [0]{\begingroup\@sanitize@url \@url }%
\providecommand \@url [1]{\endgroup\@href {#1}{\urlprefix }}%
\providecommand \urlprefix  [0]{URL }%
\providecommand \Eprint [0]{\href }%
\providecommand \doibase [0]{https://doi.org/}%
\providecommand \selectlanguage [0]{\@gobble}%
\providecommand \bibinfo  [0]{\@secondoftwo}%
\providecommand \bibfield  [0]{\@secondoftwo}%
\providecommand \translation [1]{[#1]}%
\providecommand \BibitemOpen [0]{}%
\providecommand \bibitemStop [0]{}%
\providecommand \bibitemNoStop [0]{.\EOS\space}%
\providecommand \EOS [0]{\spacefactor3000\relax}%
\providecommand \BibitemShut  [1]{\csname bibitem#1\endcsname}%
\let\auto@bib@innerbib\@empty
\bibitem [{\citenamefont {Cevallos}\ \emph {et~al.}(2018)\citenamefont
  {Cevallos}, \citenamefont {Stolze}, \citenamefont {Kong},\ and\ \citenamefont
  {Cava}}]{Cava2018}%
  \BibitemOpen
  \bibfield  {author} {\bibinfo {author} {\bibfnamefont {F.~A.}\ \bibnamefont
  {Cevallos}}, \bibinfo {author} {\bibfnamefont {K.}~\bibnamefont {Stolze}},
  \bibinfo {author} {\bibfnamefont {T.}~\bibnamefont {Kong}},\ and\ \bibinfo
  {author} {\bibfnamefont {R.~J.}\ \bibnamefont {Cava}},\ }\href
  {https://doi.org/https://doi.org/10.1016/j.materresbull.2018.04.042}
  {\bibfield  {journal} {\bibinfo  {journal} {Mater. Res. Bull.}\ }\textbf
  {\bibinfo {volume} {105}},\ \bibinfo {pages} {154} (\bibinfo {year}
  {2018})}\BibitemShut {NoStop}%
\bibitem [{\citenamefont {Shen}\ \emph {et~al.}(2019)\citenamefont {Shen},
  \citenamefont {Liu}, \citenamefont {Qin}, \citenamefont {Shen}, \citenamefont
  {Li}, \citenamefont {Bewley}, \citenamefont {Schneidewind}, \citenamefont
  {Chen},\ and\ \citenamefont {Zhao}}]{Shen2019}%
  \BibitemOpen
  \bibfield  {author} {\bibinfo {author} {\bibfnamefont {Y.}~\bibnamefont
  {Shen}}, \bibinfo {author} {\bibfnamefont {C.}~\bibnamefont {Liu}}, \bibinfo
  {author} {\bibfnamefont {Y.}~\bibnamefont {Qin}}, \bibinfo {author}
  {\bibfnamefont {S.}~\bibnamefont {Shen}}, \bibinfo {author} {\bibfnamefont
  {Y.-D.}\ \bibnamefont {Li}}, \bibinfo {author} {\bibfnamefont
  {R.}~\bibnamefont {Bewley}}, \bibinfo {author} {\bibfnamefont
  {A.}~\bibnamefont {Schneidewind}}, \bibinfo {author} {\bibfnamefont
  {G.}~\bibnamefont {Chen}},\ and\ \bibinfo {author} {\bibfnamefont
  {J.}~\bibnamefont {Zhao}},\ }\href
  {https://doi.org/10.1038/s41467-019-12410-3} {\bibfield  {journal} {\bibinfo
  {journal} {Nat. Commun.}\ }\textbf {\bibinfo {volume} {10}},\ \bibinfo
  {pages} {4530} (\bibinfo {year} {2019})}\BibitemShut {NoStop}%
\bibitem [{\citenamefont {Li}\ \emph {et~al.}(2020{\natexlab{a}})\citenamefont
  {Li}, \citenamefont {Bachus}, \citenamefont {Deng}, \citenamefont {Schmidt},
  \citenamefont {Thoma}, \citenamefont {Hutanu}, \citenamefont {Tokiwa},
  \citenamefont {Tsirlin},\ and\ \citenamefont {Gegenwart}}]{Li2020}%
  \BibitemOpen
  \bibfield  {author} {\bibinfo {author} {\bibfnamefont {Y.}~\bibnamefont
  {Li}}, \bibinfo {author} {\bibfnamefont {S.}~\bibnamefont {Bachus}}, \bibinfo
  {author} {\bibfnamefont {H.}~\bibnamefont {Deng}}, \bibinfo {author}
  {\bibfnamefont {W.}~\bibnamefont {Schmidt}}, \bibinfo {author} {\bibfnamefont
  {H.}~\bibnamefont {Thoma}}, \bibinfo {author} {\bibfnamefont
  {V.}~\bibnamefont {Hutanu}}, \bibinfo {author} {\bibfnamefont
  {Y.}~\bibnamefont {Tokiwa}}, \bibinfo {author} {\bibfnamefont {A.~A.}\
  \bibnamefont {Tsirlin}},\ and\ \bibinfo {author} {\bibfnamefont
  {P.}~\bibnamefont {Gegenwart}},\ }\href
  {https://doi.org/10.1103/PhysRevX.10.011007} {\bibfield  {journal} {\bibinfo
  {journal} {Phys. Rev. X}\ }\textbf {\bibinfo {volume} {10}},\ \bibinfo
  {pages} {011007} (\bibinfo {year} {2020}{\natexlab{a}})}\BibitemShut
  {NoStop}%
\bibitem [{\citenamefont {Hu}\ \emph {et~al.}(2020)\citenamefont {Hu},
  \citenamefont {Ma}, \citenamefont {Liao}, \citenamefont {Li}, \citenamefont
  {Ma}, \citenamefont {Cui}, \citenamefont {Shangguan}, \citenamefont {Huang},
  \citenamefont {Qi}, \citenamefont {Li}, \citenamefont {Meng}, \citenamefont
  {Wen},\ and\ \citenamefont {Yu}}]{ZHu2020}%
  \BibitemOpen
  \bibfield  {author} {\bibinfo {author} {\bibfnamefont {Z.}~\bibnamefont
  {Hu}}, \bibinfo {author} {\bibfnamefont {Z.}~\bibnamefont {Ma}}, \bibinfo
  {author} {\bibfnamefont {Y.-D.}\ \bibnamefont {Liao}}, \bibinfo {author}
  {\bibfnamefont {H.}~\bibnamefont {Li}}, \bibinfo {author} {\bibfnamefont
  {C.}~\bibnamefont {Ma}}, \bibinfo {author} {\bibfnamefont {Y.}~\bibnamefont
  {Cui}}, \bibinfo {author} {\bibfnamefont {Y.}~\bibnamefont {Shangguan}},
  \bibinfo {author} {\bibfnamefont {Z.}~\bibnamefont {Huang}}, \bibinfo
  {author} {\bibfnamefont {Y.}~\bibnamefont {Qi}}, \bibinfo {author}
  {\bibfnamefont {W.}~\bibnamefont {Li}}, \bibinfo {author} {\bibfnamefont
  {Z.~Y.}\ \bibnamefont {Meng}}, \bibinfo {author} {\bibfnamefont
  {J.}~\bibnamefont {Wen}},\ and\ \bibinfo {author} {\bibfnamefont
  {W.}~\bibnamefont {Yu}},\ }\href {https://doi.org/10.1038/s41467-020-19380-x}
  {\bibfield  {journal} {\bibinfo  {journal} {Nature Communications}\ }\textbf
  {\bibinfo {volume} {11}},\ \bibinfo {pages} {5631} (\bibinfo {year}
  {2020})}\BibitemShut {NoStop}%
\bibitem [{\citenamefont {Li}\ \emph {et~al.}(2020{\natexlab{b}})\citenamefont
  {Li}, \citenamefont {Liao}, \citenamefont {Chen}, \citenamefont {Zeng},
  \citenamefont {Sheng}, \citenamefont {Qi}, \citenamefont {Meng},\ and\
  \citenamefont {Li}}]{Lih2020}%
  \BibitemOpen
  \bibfield  {author} {\bibinfo {author} {\bibfnamefont {H.}~\bibnamefont
  {Li}}, \bibinfo {author} {\bibfnamefont {Y.~D.}\ \bibnamefont {Liao}},
  \bibinfo {author} {\bibfnamefont {B.-B.}\ \bibnamefont {Chen}}, \bibinfo
  {author} {\bibfnamefont {X.-T.}\ \bibnamefont {Zeng}}, \bibinfo {author}
  {\bibfnamefont {X.-L.}\ \bibnamefont {Sheng}}, \bibinfo {author}
  {\bibfnamefont {Y.}~\bibnamefont {Qi}}, \bibinfo {author} {\bibfnamefont
  {Z.~Y.}\ \bibnamefont {Meng}},\ and\ \bibinfo {author} {\bibfnamefont
  {W.}~\bibnamefont {Li}},\ }\href {https://doi.org/10.1038/s41467-020-14907-8}
  {\bibfield  {journal} {\bibinfo  {journal} {Nat. Commun.}\ }\textbf {\bibinfo
  {volume} {11}},\ \bibinfo {pages} {1111} (\bibinfo {year}
  {2020}{\natexlab{b}})}\BibitemShut {NoStop}%
\bibitem [{\citenamefont {Liu}\ \emph {et~al.}(2020)\citenamefont {Liu},
  \citenamefont {Huang},\ and\ \citenamefont {Chen}}]{Liu2020Intrinsic}%
  \BibitemOpen
  \bibfield  {author} {\bibinfo {author} {\bibfnamefont {C.}~\bibnamefont
  {Liu}}, \bibinfo {author} {\bibfnamefont {C.-J.}\ \bibnamefont {Huang}},\
  and\ \bibinfo {author} {\bibfnamefont {G.}~\bibnamefont {Chen}},\ }\href
  {https://doi.org/10.1103/PhysRevResearch.2.043013} {\bibfield  {journal}
  {\bibinfo  {journal} {Phys. Rev. Research}\ }\textbf {\bibinfo {volume}
  {2}},\ \bibinfo {pages} {043013} (\bibinfo {year} {2020})}\BibitemShut
  {NoStop}%
\bibitem [{\citenamefont {Moessner}\ and\ \citenamefont
  {Sondhi}(2001)}]{Moessner2001}%
  \BibitemOpen
  \bibfield  {author} {\bibinfo {author} {\bibfnamefont {R.}~\bibnamefont
  {Moessner}}\ and\ \bibinfo {author} {\bibfnamefont {S.~L.}\ \bibnamefont
  {Sondhi}},\ }\href {https://doi.org/10.1103/physrevlett.86.1881} {\bibfield
  {journal} {\bibinfo  {journal} {Phys. Rev. Lett.}\ }\textbf {\bibinfo
  {volume} {86}},\ \bibinfo {pages} {1881} (\bibinfo {year}
  {2001})}\BibitemShut {NoStop}%
\bibitem [{\citenamefont {Isakov}\ and\ \citenamefont
  {Moessner}(2003)}]{Isakov2003}%
  \BibitemOpen
  \bibfield  {author} {\bibinfo {author} {\bibfnamefont {S.~V.}\ \bibnamefont
  {Isakov}}\ and\ \bibinfo {author} {\bibfnamefont {R.}~\bibnamefont
  {Moessner}},\ }\href {https://doi.org/10.1103/PhysRevB.68.104409} {\bibfield
  {journal} {\bibinfo  {journal} {Phys. Rev. B}\ }\textbf {\bibinfo {volume}
  {68}},\ \bibinfo {pages} {104409} (\bibinfo {year} {2003})}\BibitemShut
  {NoStop}%
\bibitem [{\citenamefont {Wang}\ \emph {et~al.}(2017)\citenamefont {Wang},
  \citenamefont {Qi}, \citenamefont {Chen},\ and\ \citenamefont
  {Meng}}]{YCWang2017}%
  \BibitemOpen
  \bibfield  {author} {\bibinfo {author} {\bibfnamefont {Y.-C.}\ \bibnamefont
  {Wang}}, \bibinfo {author} {\bibfnamefont {Y.}~\bibnamefont {Qi}}, \bibinfo
  {author} {\bibfnamefont {S.}~\bibnamefont {Chen}},\ and\ \bibinfo {author}
  {\bibfnamefont {Z.~Y.}\ \bibnamefont {Meng}},\ }\href
  {https://doi.org/10.1103/PhysRevB.96.115160} {\bibfield  {journal} {\bibinfo
  {journal} {Phys. Rev. B}\ }\textbf {\bibinfo {volume} {96}},\ \bibinfo
  {pages} {115160} (\bibinfo {year} {2017})}\BibitemShut {NoStop}%
\bibitem [{\citenamefont {Dun}\ \emph {et~al.}(2020)\citenamefont {Dun},
  \citenamefont {Daum}, \citenamefont {Baral}, \citenamefont {Fischer},
  \citenamefont {Cao}, \citenamefont {Liu}, \citenamefont {Stone},
  \citenamefont {Rodriguez-Rivera}, \citenamefont {Choi}, \citenamefont
  {Huang}, \citenamefont {Zhou}, \citenamefont {Mourigal},\ and\ \citenamefont
  {Frandsen}}]{Dun2020neutron}%
  \BibitemOpen
  \bibfield  {author} {\bibinfo {author} {\bibfnamefont {Z.}~\bibnamefont
  {Dun}}, \bibinfo {author} {\bibfnamefont {M.}~\bibnamefont {Daum}}, \bibinfo
  {author} {\bibfnamefont {R.}~\bibnamefont {Baral}}, \bibinfo {author}
  {\bibfnamefont {H.~E.}\ \bibnamefont {Fischer}}, \bibinfo {author}
  {\bibfnamefont {H.}~\bibnamefont {Cao}}, \bibinfo {author} {\bibfnamefont
  {Y.}~\bibnamefont {Liu}}, \bibinfo {author} {\bibfnamefont {M.~B.}\
  \bibnamefont {Stone}}, \bibinfo {author} {\bibfnamefont {J.~A.}\ \bibnamefont
  {Rodriguez-Rivera}}, \bibinfo {author} {\bibfnamefont {E.~S.}\ \bibnamefont
  {Choi}}, \bibinfo {author} {\bibfnamefont {Q.}~\bibnamefont {Huang}},
  \bibinfo {author} {\bibfnamefont {H.}~\bibnamefont {Zhou}}, \bibinfo {author}
  {\bibfnamefont {M.}~\bibnamefont {Mourigal}},\ and\ \bibinfo {author}
  {\bibfnamefont {B.}~\bibnamefont {Frandsen}},\ }\href@noop {} {\bibinfo
  {title} {Neutron scattering investigation of proposed kosterlitz-thouless
  transitions in the triangular-lattice ising antiferromagnet tmmggao4}}
  (\bibinfo {year} {2020}),\ \Eprint {https://arxiv.org/abs/2011.00541}
  {arXiv:2011.00541 [cond-mat.str-el]} \BibitemShut {NoStop}%
\bibitem [{\citenamefont {Qin}\ \emph {et~al.}(2020)\citenamefont {Qin},
  \citenamefont {Shen}, \citenamefont {Liu}, \citenamefont {Wo}, \citenamefont
  {Gao}, \citenamefont {Feng}, \citenamefont {Zhang}, \citenamefont {Ding},
  \citenamefont {Gu}, \citenamefont {Wang}, \citenamefont {Shen}, \citenamefont
  {Walker}, \citenamefont {Bewley}, \citenamefont {Xu}, \citenamefont {Boehm},
  \citenamefont {Steffens}, \citenamefont {Ohira-Kawamura}, \citenamefont
  {Murai}, \citenamefont {Schneidewind}, \citenamefont {Tong}, \citenamefont
  {Chen},\ and\ \citenamefont {Zhao}}]{Qin2020fieldtuned}%
  \BibitemOpen
  \bibfield  {author} {\bibinfo {author} {\bibfnamefont {Y.}~\bibnamefont
  {Qin}}, \bibinfo {author} {\bibfnamefont {Y.}~\bibnamefont {Shen}}, \bibinfo
  {author} {\bibfnamefont {C.}~\bibnamefont {Liu}}, \bibinfo {author}
  {\bibfnamefont {H.}~\bibnamefont {Wo}}, \bibinfo {author} {\bibfnamefont
  {Y.}~\bibnamefont {Gao}}, \bibinfo {author} {\bibfnamefont {Y.}~\bibnamefont
  {Feng}}, \bibinfo {author} {\bibfnamefont {X.}~\bibnamefont {Zhang}},
  \bibinfo {author} {\bibfnamefont {G.}~\bibnamefont {Ding}}, \bibinfo {author}
  {\bibfnamefont {Y.}~\bibnamefont {Gu}}, \bibinfo {author} {\bibfnamefont
  {Q.}~\bibnamefont {Wang}}, \bibinfo {author} {\bibfnamefont {S.}~\bibnamefont
  {Shen}}, \bibinfo {author} {\bibfnamefont {H.~C.}\ \bibnamefont {Walker}},
  \bibinfo {author} {\bibfnamefont {R.}~\bibnamefont {Bewley}}, \bibinfo
  {author} {\bibfnamefont {J.}~\bibnamefont {Xu}}, \bibinfo {author}
  {\bibfnamefont {M.}~\bibnamefont {Boehm}}, \bibinfo {author} {\bibfnamefont
  {P.}~\bibnamefont {Steffens}}, \bibinfo {author} {\bibfnamefont
  {S.}~\bibnamefont {Ohira-Kawamura}}, \bibinfo {author} {\bibfnamefont
  {N.}~\bibnamefont {Murai}}, \bibinfo {author} {\bibfnamefont
  {A.}~\bibnamefont {Schneidewind}}, \bibinfo {author} {\bibfnamefont
  {X.}~\bibnamefont {Tong}}, \bibinfo {author} {\bibfnamefont {G.}~\bibnamefont
  {Chen}},\ and\ \bibinfo {author} {\bibfnamefont {J.}~\bibnamefont {Zhao}},\
  }\href@noop {} {\bibinfo {title} {Field-tuned quantum effects in a
  triangular-lattice ising magnet}} (\bibinfo {year} {2020}),\ \Eprint
  {https://arxiv.org/abs/2011.09376} {arXiv:2011.09376 [cond-mat.str-el]}
  \BibitemShut {NoStop}%
\bibitem [{\citenamefont {Huang}\ \emph {et~al.}(2020)\citenamefont {Huang},
  \citenamefont {Wang}, \citenamefont {Wang},\ and\ \citenamefont
  {Chen}}]{Huang2020emergent}%
  \BibitemOpen
  \bibfield  {author} {\bibinfo {author} {\bibfnamefont {C.-J.}\ \bibnamefont
  {Huang}}, \bibinfo {author} {\bibfnamefont {X.}~\bibnamefont {Wang}},
  \bibinfo {author} {\bibfnamefont {Z.}~\bibnamefont {Wang}},\ and\ \bibinfo
  {author} {\bibfnamefont {G.}~\bibnamefont {Chen}},\ }\href@noop {} {\bibinfo
  {title} {Emergent halperin-saslow mode and gauge glass in quantum ising
  magnet tmmggao$_4$}} (\bibinfo {year} {2020}),\ \Eprint
  {https://arxiv.org/abs/2011.05919} {arXiv:2011.05919 [cond-mat.str-el]}
  \BibitemShut {NoStop}%
\bibitem [{\citenamefont {{Zhou}}\ \emph
  {et~al.}(2020{\natexlab{a}})\citenamefont {{Zhou}}, \citenamefont {{Liu}},
  \citenamefont {{Yan}}, \citenamefont {{Chen}},\ and\ \citenamefont
  {{Zhang}}}]{ZhengZhou202005}%
  \BibitemOpen
  \bibfield  {author} {\bibinfo {author} {\bibfnamefont {Z.}~\bibnamefont
  {{Zhou}}}, \bibinfo {author} {\bibfnamefont {D.-X.}\ \bibnamefont {{Liu}}},
  \bibinfo {author} {\bibfnamefont {Z.}~\bibnamefont {{Yan}}}, \bibinfo
  {author} {\bibfnamefont {Y.}~\bibnamefont {{Chen}}},\ and\ \bibinfo {author}
  {\bibfnamefont {X.-F.}\ \bibnamefont {{Zhang}}},\ }\href@noop {} {\bibfield
  {journal} {\bibinfo  {journal} {arXiv e-prints}\ ,\ \bibinfo {eid}
  {arXiv:2005.11133}} (\bibinfo {year} {2020}{\natexlab{a}})},\ \Eprint
  {https://arxiv.org/abs/2005.11133} {arXiv:2005.11133 [cond-mat.str-el]}
  \BibitemShut {NoStop}%
\bibitem [{\citenamefont {{Zhou}}\ \emph
  {et~al.}(2020{\natexlab{b}})\citenamefont {{Zhou}}, \citenamefont {{Liu}},
  \citenamefont {{Yan}}, \citenamefont {{Chen}},\ and\ \citenamefont
  {{Zhang}}}]{ZhengZhou202010}%
  \BibitemOpen
  \bibfield  {author} {\bibinfo {author} {\bibfnamefont {Z.}~\bibnamefont
  {{Zhou}}}, \bibinfo {author} {\bibfnamefont {C.-L.}\ \bibnamefont {{Liu}}},
  \bibinfo {author} {\bibfnamefont {Z.}~\bibnamefont {{Yan}}}, \bibinfo
  {author} {\bibfnamefont {Y.}~\bibnamefont {{Chen}}},\ and\ \bibinfo {author}
  {\bibfnamefont {X.-F.}\ \bibnamefont {{Zhang}}},\ }\href@noop {} {\bibfield
  {journal} {\bibinfo  {journal} {arXiv e-prints}\ ,\ \bibinfo {eid}
  {arXiv:2010.01750}} (\bibinfo {year} {2020}{\natexlab{b}})},\ \Eprint
  {https://arxiv.org/abs/2010.01750} {arXiv:2010.01750 [cond-mat.str-el]}
  \BibitemShut {NoStop}%
\bibitem [{\citenamefont {Sandvik}(2010)}]{Sandvik2010}%
  \BibitemOpen
  \bibfield  {author} {\bibinfo {author} {\bibfnamefont {A.~W.}\ \bibnamefont
  {Sandvik}},\ }\href {https://doi.org/10.1063/1.3518900} {\bibfield  {journal}
  {\bibinfo  {journal} {AIP Conference Proceedings}\ }\textbf {\bibinfo
  {volume} {1297}},\ \bibinfo {pages} {135} (\bibinfo {year}
  {2010})}\BibitemShut {NoStop}%
\bibitem [{\citenamefont {Sandvik}(2003)}]{sandvikTFIM}%
  \BibitemOpen
  \bibfield  {author} {\bibinfo {author} {\bibfnamefont {A.~W.}\ \bibnamefont
  {Sandvik}},\ }\href
  {https://journals.aps.org/pre/abstract/10.1103/PhysRevE.68.056701} {\bibfield
   {journal} {\bibinfo  {journal} {Physical Review E}\ }\textbf {\bibinfo
  {volume} {68}},\ \bibinfo {pages} {056701} (\bibinfo {year}
  {2003})}\BibitemShut {NoStop}%
\bibitem [{\citenamefont {Melko}(2013)}]{melko2013stochastic}%
  \BibitemOpen
  \bibfield  {author} {\bibinfo {author} {\bibfnamefont {R.~G.}\ \bibnamefont
  {Melko}},\ }in\ \href
  {https://link.springer.com/chapter/10.1007/978-3-642-35106-8_7} {\emph
  {\bibinfo {booktitle} {Strongly Correlated Systems}}}\ (\bibinfo  {publisher}
  {Springer},\ \bibinfo {year} {2013})\ pp.\ \bibinfo {pages}
  {185--206}\BibitemShut {NoStop}%
\bibitem [{\citenamefont {{Zhao}}\ \emph {et~al.}(2020)\citenamefont {{Zhao}},
  \citenamefont {{Yan}}, \citenamefont {{Cheng}},\ and\ \citenamefont
  {{Meng}}}]{JRZhao2020}%
  \BibitemOpen
  \bibfield  {author} {\bibinfo {author} {\bibfnamefont {J.}~\bibnamefont
  {{Zhao}}}, \bibinfo {author} {\bibfnamefont {Z.}~\bibnamefont {{Yan}}},
  \bibinfo {author} {\bibfnamefont {M.}~\bibnamefont {{Cheng}}},\ and\ \bibinfo
  {author} {\bibfnamefont {Z.~Y.}\ \bibnamefont {{Meng}}},\ }\href@noop {}
  {\bibfield  {journal} {\bibinfo  {journal} {arXiv e-prints}\ ,\ \bibinfo
  {eid} {arXiv:2011.12543}} (\bibinfo {year} {2020})},\ \Eprint
  {https://arxiv.org/abs/2011.12543} {arXiv:2011.12543 [cond-mat.str-el]}
  \BibitemShut {NoStop}%
\bibitem [{\citenamefont {Kato}\ and\ \citenamefont
  {Misawa}(2015)}]{takahiro2015}%
  \BibitemOpen
  \bibfield  {author} {\bibinfo {author} {\bibfnamefont {Y.}~\bibnamefont
  {Kato}}\ and\ \bibinfo {author} {\bibfnamefont {T.}~\bibnamefont {Misawa}},\
  }\href {https://doi.org/10.1103/PhysRevB.92.174419} {\bibfield  {journal}
  {\bibinfo  {journal} {Phys. Rev. B}\ }\textbf {\bibinfo {volume} {92}},\
  \bibinfo {pages} {174419} (\bibinfo {year} {2015})}\BibitemShut {NoStop}%
\bibitem [{\citenamefont {Liu}\ \emph {et~al.}(2021)\citenamefont {Liu},
  \citenamefont {Meng},\ and\ \citenamefont {Yin}}]{yuzhi2020}%
  \BibitemOpen
  \bibfield  {author} {\bibinfo {author} {\bibfnamefont {Y.}~\bibnamefont
  {Liu}}, \bibinfo {author} {\bibfnamefont {Z.~Y.}\ \bibnamefont {Meng}},\ and\
  \bibinfo {author} {\bibfnamefont {S.}~\bibnamefont {Yin}},\ }\href
  {https://doi.org/10.1103/PhysRevB.103.075147} {\bibfield  {journal} {\bibinfo
   {journal} {Phys. Rev. B}\ }\textbf {\bibinfo {volume} {103}},\ \bibinfo
  {pages} {075147} (\bibinfo {year} {2021})}\BibitemShut {NoStop}%
\bibitem [{\citenamefont {Binder}(1981{\natexlab{a}})}]{Binder-1981a}%
  \BibitemOpen
  \bibfield  {author} {\bibinfo {author} {\bibfnamefont {K.}~\bibnamefont
  {Binder}},\ }\href {https://doi.org/10.1103/PhysRevLett.47.693} {\bibfield
  {journal} {\bibinfo  {journal} {Phys. Rev. Lett.}\ }\textbf {\bibinfo
  {volume} {47}},\ \bibinfo {pages} {693} (\bibinfo {year}
  {1981}{\natexlab{a}})}\BibitemShut {NoStop}%
\bibitem [{\citenamefont {Binder}(1981{\natexlab{b}})}]{Binder-1981b}%
  \BibitemOpen
  \bibfield  {author} {\bibinfo {author} {\bibfnamefont {K.}~\bibnamefont
  {Binder}},\ }\href {https://doi.org/10.1007/BF01293604} {\bibfield  {journal}
  {\bibinfo  {journal} {Z. Phys. B}\ }\textbf {\bibinfo {volume} {43}},\
  \bibinfo {pages} {119} (\bibinfo {year} {1981}{\natexlab{b}})}\BibitemShut
  {NoStop}%
\bibitem [{\citenamefont {Binder}\ and\ \citenamefont
  {Landau}(1984)}]{Binder1984}%
  \BibitemOpen
  \bibfield  {author} {\bibinfo {author} {\bibfnamefont {K.}~\bibnamefont
  {Binder}}\ and\ \bibinfo {author} {\bibfnamefont {D.~P.}\ \bibnamefont
  {Landau}},\ }\href {https://doi.org/10.1103/PhysRevB.30.1477} {\bibfield
  {journal} {\bibinfo  {journal} {Phys. Rev. B}\ }\textbf {\bibinfo {volume}
  {30}},\ \bibinfo {pages} {1477} (\bibinfo {year} {1984})}\BibitemShut
  {NoStop}%
\bibitem [{\citenamefont {Samajdar}\ \emph {et~al.}(2021)\citenamefont
  {Samajdar}, \citenamefont {Ho}, \citenamefont {Pichler}, \citenamefont
  {Lukin},\ and\ \citenamefont {Sachdev}}]{Rhine2021}%
  \BibitemOpen
  \bibfield  {author} {\bibinfo {author} {\bibfnamefont {R.}~\bibnamefont
  {Samajdar}}, \bibinfo {author} {\bibfnamefont {W.~W.}\ \bibnamefont {Ho}},
  \bibinfo {author} {\bibfnamefont {H.}~\bibnamefont {Pichler}}, \bibinfo
  {author} {\bibfnamefont {M.~D.}\ \bibnamefont {Lukin}},\ and\ \bibinfo
  {author} {\bibfnamefont {S.}~\bibnamefont {Sachdev}},\ }\bibfield  {journal}
  {\bibinfo  {journal} {Proceedings of the National Academy of Sciences}\
  }\textbf {\bibinfo {volume} {118}},\ \href
  {https://doi.org/10.1073/pnas.2015785118} {10.1073/pnas.2015785118} (\bibinfo
  {year} {2021})\BibitemShut {NoStop}%
\bibitem [{\citenamefont {{Ebadi}}\ \emph {et~al.}(2020)\citenamefont
  {{Ebadi}}, \citenamefont {{Wang}}, \citenamefont {{Levine}}, \citenamefont
  {{Keesling}}, \citenamefont {{Semeghini}}, \citenamefont {{Omran}},
  \citenamefont {{Bluvstein}}, \citenamefont {{Samajdar}}, \citenamefont
  {{Pichler}}, \citenamefont {{Ho}}, \citenamefont {{Choi}}, \citenamefont
  {{Sachdev}}, \citenamefont {{Greiner}}, \citenamefont {{Vuletic}},\ and\
  \citenamefont {{Lukin}}}]{Ebadi2020}%
  \BibitemOpen
  \bibfield  {author} {\bibinfo {author} {\bibfnamefont {S.}~\bibnamefont
  {{Ebadi}}}, \bibinfo {author} {\bibfnamefont {T.~T.}\ \bibnamefont {{Wang}}},
  \bibinfo {author} {\bibfnamefont {H.}~\bibnamefont {{Levine}}}, \bibinfo
  {author} {\bibfnamefont {A.}~\bibnamefont {{Keesling}}}, \bibinfo {author}
  {\bibfnamefont {G.}~\bibnamefont {{Semeghini}}}, \bibinfo {author}
  {\bibfnamefont {A.}~\bibnamefont {{Omran}}}, \bibinfo {author} {\bibfnamefont
  {D.}~\bibnamefont {{Bluvstein}}}, \bibinfo {author} {\bibfnamefont
  {R.}~\bibnamefont {{Samajdar}}}, \bibinfo {author} {\bibfnamefont
  {H.}~\bibnamefont {{Pichler}}}, \bibinfo {author} {\bibfnamefont {W.~W.}\
  \bibnamefont {{Ho}}}, \bibinfo {author} {\bibfnamefont {S.}~\bibnamefont
  {{Choi}}}, \bibinfo {author} {\bibfnamefont {S.}~\bibnamefont {{Sachdev}}},
  \bibinfo {author} {\bibfnamefont {M.}~\bibnamefont {{Greiner}}}, \bibinfo
  {author} {\bibfnamefont {V.}~\bibnamefont {{Vuletic}}},\ and\ \bibinfo
  {author} {\bibfnamefont {M.~D.}\ \bibnamefont {{Lukin}}},\ }\href@noop {}
  {\bibfield  {journal} {\bibinfo  {journal} {arXiv e-prints}\ ,\ \bibinfo
  {eid} {arXiv:2012.12281}} (\bibinfo {year} {2020})},\ \Eprint
  {https://arxiv.org/abs/2012.12281} {arXiv:2012.12281 [quant-ph]} \BibitemShut
  {NoStop}%
\bibitem [{\citenamefont {Scholl}\ \emph {et~al.}(2020)\citenamefont {Scholl},
  \citenamefont {Schuler}, \citenamefont {Williams}, \citenamefont
  {Eberharter}, \citenamefont {Barredo}, \citenamefont {Schymik}, \citenamefont
  {Lienhard}, \citenamefont {Henry}, \citenamefont {Lang}, \citenamefont
  {Lahaye}, \citenamefont {Läuchli},\ and\ \citenamefont
  {Browaeys}}]{scholl2020programmable}%
  \BibitemOpen
  \bibfield  {author} {\bibinfo {author} {\bibfnamefont {P.}~\bibnamefont
  {Scholl}}, \bibinfo {author} {\bibfnamefont {M.}~\bibnamefont {Schuler}},
  \bibinfo {author} {\bibfnamefont {H.~J.}\ \bibnamefont {Williams}}, \bibinfo
  {author} {\bibfnamefont {A.~A.}\ \bibnamefont {Eberharter}}, \bibinfo
  {author} {\bibfnamefont {D.}~\bibnamefont {Barredo}}, \bibinfo {author}
  {\bibfnamefont {K.-N.}\ \bibnamefont {Schymik}}, \bibinfo {author}
  {\bibfnamefont {V.}~\bibnamefont {Lienhard}}, \bibinfo {author}
  {\bibfnamefont {L.-P.}\ \bibnamefont {Henry}}, \bibinfo {author}
  {\bibfnamefont {T.~C.}\ \bibnamefont {Lang}}, \bibinfo {author}
  {\bibfnamefont {T.}~\bibnamefont {Lahaye}}, \bibinfo {author} {\bibfnamefont
  {A.~M.}\ \bibnamefont {Läuchli}},\ and\ \bibinfo {author} {\bibfnamefont
  {A.}~\bibnamefont {Browaeys}},\ }\href@noop {} {\bibinfo {title}
  {Programmable quantum simulation of 2d antiferromagnets with hundreds of
  rydberg atoms}} (\bibinfo {year} {2020}),\ \Eprint
  {https://arxiv.org/abs/2012.12268} {arXiv:2012.12268 [quant-ph]} \BibitemShut
  {NoStop}%
\bibitem [{\citenamefont {King}\ \emph {et~al.}(2018)\citenamefont {King},
  \citenamefont {Carrasquilla}, \citenamefont {Raymond}, \citenamefont
  {Ozfidan}, \citenamefont {Andriyash}, \citenamefont {Berkley}, \citenamefont
  {Reis}, \citenamefont {Lanting}, \citenamefont {Harris}, \citenamefont
  {Altomare}, \citenamefont {Boothby}, \citenamefont {Bunyk}, \citenamefont
  {Enderud}, \citenamefont {Fr{\'{e}}chette}, \citenamefont {Hoskinson},
  \citenamefont {Ladizinsky}, \citenamefont {Oh}, \citenamefont
  {Poulin-Lamarre}, \citenamefont {Rich}, \citenamefont {Sato}, \citenamefont
  {Smirnov}, \citenamefont {Swenson}, \citenamefont {Volkmann}, \citenamefont
  {Whittaker}, \citenamefont {Yao}, \citenamefont {Ladizinsky}, \citenamefont
  {Johnson}, \citenamefont {Hilton},\ and\ \citenamefont
  {Amin}}]{King2018topology}%
  \BibitemOpen
  \bibfield  {author} {\bibinfo {author} {\bibfnamefont {A.~D.}\ \bibnamefont
  {King}}, \bibinfo {author} {\bibfnamefont {J.}~\bibnamefont {Carrasquilla}},
  \bibinfo {author} {\bibfnamefont {J.}~\bibnamefont {Raymond}}, \bibinfo
  {author} {\bibfnamefont {I.}~\bibnamefont {Ozfidan}}, \bibinfo {author}
  {\bibfnamefont {E.}~\bibnamefont {Andriyash}}, \bibinfo {author}
  {\bibfnamefont {A.}~\bibnamefont {Berkley}}, \bibinfo {author} {\bibfnamefont
  {M.}~\bibnamefont {Reis}}, \bibinfo {author} {\bibfnamefont {T.}~\bibnamefont
  {Lanting}}, \bibinfo {author} {\bibfnamefont {R.}~\bibnamefont {Harris}},
  \bibinfo {author} {\bibfnamefont {F.}~\bibnamefont {Altomare}}, \bibinfo
  {author} {\bibfnamefont {K.}~\bibnamefont {Boothby}}, \bibinfo {author}
  {\bibfnamefont {P.~I.}\ \bibnamefont {Bunyk}}, \bibinfo {author}
  {\bibfnamefont {C.}~\bibnamefont {Enderud}}, \bibinfo {author} {\bibfnamefont
  {A.}~\bibnamefont {Fr{\'{e}}chette}}, \bibinfo {author} {\bibfnamefont
  {E.}~\bibnamefont {Hoskinson}}, \bibinfo {author} {\bibfnamefont
  {N.}~\bibnamefont {Ladizinsky}}, \bibinfo {author} {\bibfnamefont
  {T.}~\bibnamefont {Oh}}, \bibinfo {author} {\bibfnamefont {G.}~\bibnamefont
  {Poulin-Lamarre}}, \bibinfo {author} {\bibfnamefont {C.}~\bibnamefont
  {Rich}}, \bibinfo {author} {\bibfnamefont {Y.}~\bibnamefont {Sato}}, \bibinfo
  {author} {\bibfnamefont {A.~Y.}\ \bibnamefont {Smirnov}}, \bibinfo {author}
  {\bibfnamefont {L.~J.}\ \bibnamefont {Swenson}}, \bibinfo {author}
  {\bibfnamefont {M.~H.}\ \bibnamefont {Volkmann}}, \bibinfo {author}
  {\bibfnamefont {J.}~\bibnamefont {Whittaker}}, \bibinfo {author}
  {\bibfnamefont {J.}~\bibnamefont {Yao}}, \bibinfo {author} {\bibfnamefont
  {E.}~\bibnamefont {Ladizinsky}}, \bibinfo {author} {\bibfnamefont {M.~W.}\
  \bibnamefont {Johnson}}, \bibinfo {author} {\bibfnamefont {J.}~\bibnamefont
  {Hilton}},\ and\ \bibinfo {author} {\bibfnamefont {M.~H.}\ \bibnamefont
  {Amin}},\ }\href {https://doi.org/10.1038/s41586-018-0410-x} {\bibfield
  {journal} {\bibinfo  {journal} {Nature}\ }\textbf {\bibinfo {volume} {560}},\
  \bibinfo {pages} {456} (\bibinfo {year} {2018})}\BibitemShut {NoStop}%
\bibitem [{\citenamefont {{King}}\ \emph {et~al.}(2019)\citenamefont {{King}},
  \citenamefont {{Raymond}}, \citenamefont {{Lanting}}, \citenamefont
  {{Isakov}}, \citenamefont {{Mohseni}}, \citenamefont {{Poulin-Lamarre}},
  \citenamefont {{Ejtemaee}}, \citenamefont {{Bernoudy}}, \citenamefont
  {{Ozfidan}}, \citenamefont {{Smirnov}}, \citenamefont {{Reis}}, \citenamefont
  {{Altomare}}, \citenamefont {{Babcock}}, \citenamefont {{Baron}},
  \citenamefont {{Berkley}}, \citenamefont {{Boothby}}, \citenamefont
  {{Bunyk}}, \citenamefont {{Christiani}}, \citenamefont {{Enderud}},
  \citenamefont {{Evert}}, \citenamefont {{Harris}}, \citenamefont
  {{Hoskinson}}, \citenamefont {{Huang}}, \citenamefont {{Jooya}},
  \citenamefont {{Khodabandelou}}, \citenamefont {{Ladizinsky}}, \citenamefont
  {{Li}}, \citenamefont {{Lott}}, \citenamefont {{MacDonald}}, \citenamefont
  {{Marsden}}, \citenamefont {{Marsden}}, \citenamefont {{Medina}},
  \citenamefont {{Molavi}}, \citenamefont {{Neufeld}}, \citenamefont
  {{Norouzpour}}, \citenamefont {{Oh}}, \citenamefont {{Pavlov}}, \citenamefont
  {{Perminov}}, \citenamefont {{Prescott}}, \citenamefont {{Rich}},
  \citenamefont {{Sato}}, \citenamefont {{Sheldan}}, \citenamefont
  {{Sterling}}, \citenamefont {{Swenson}}, \citenamefont {{Tsai}},
  \citenamefont {{Volkmann}}, \citenamefont {{Whittaker}}, \citenamefont
  {{Wilkinson}}, \citenamefont {{Yao}}, \citenamefont {{Neven}}, \citenamefont
  {{Hilton}}, \citenamefont {{Ladizinsky}}, \citenamefont {{Johnson}},\ and\
  \citenamefont {{Amin}}}]{King2019scalingadv}%
  \BibitemOpen
  \bibfield  {author} {\bibinfo {author} {\bibfnamefont {A.~D.}\ \bibnamefont
  {{King}}}, \bibinfo {author} {\bibfnamefont {J.}~\bibnamefont {{Raymond}}},
  \bibinfo {author} {\bibfnamefont {T.}~\bibnamefont {{Lanting}}}, \bibinfo
  {author} {\bibfnamefont {S.~V.}\ \bibnamefont {{Isakov}}}, \bibinfo {author}
  {\bibfnamefont {M.}~\bibnamefont {{Mohseni}}}, \bibinfo {author}
  {\bibfnamefont {G.}~\bibnamefont {{Poulin-Lamarre}}}, \bibinfo {author}
  {\bibfnamefont {S.}~\bibnamefont {{Ejtemaee}}}, \bibinfo {author}
  {\bibfnamefont {W.}~\bibnamefont {{Bernoudy}}}, \bibinfo {author}
  {\bibfnamefont {I.}~\bibnamefont {{Ozfidan}}}, \bibinfo {author}
  {\bibfnamefont {A.~Y.}\ \bibnamefont {{Smirnov}}}, \bibinfo {author}
  {\bibfnamefont {M.}~\bibnamefont {{Reis}}}, \bibinfo {author} {\bibfnamefont
  {F.}~\bibnamefont {{Altomare}}}, \bibinfo {author} {\bibfnamefont
  {M.}~\bibnamefont {{Babcock}}}, \bibinfo {author} {\bibfnamefont
  {C.}~\bibnamefont {{Baron}}}, \bibinfo {author} {\bibfnamefont {A.~J.}\
  \bibnamefont {{Berkley}}}, \bibinfo {author} {\bibfnamefont {K.}~\bibnamefont
  {{Boothby}}}, \bibinfo {author} {\bibfnamefont {P.~I.}\ \bibnamefont
  {{Bunyk}}}, \bibinfo {author} {\bibfnamefont {H.}~\bibnamefont
  {{Christiani}}}, \bibinfo {author} {\bibfnamefont {C.}~\bibnamefont
  {{Enderud}}}, \bibinfo {author} {\bibfnamefont {B.}~\bibnamefont {{Evert}}},
  \bibinfo {author} {\bibfnamefont {R.}~\bibnamefont {{Harris}}}, \bibinfo
  {author} {\bibfnamefont {E.}~\bibnamefont {{Hoskinson}}}, \bibinfo {author}
  {\bibfnamefont {S.}~\bibnamefont {{Huang}}}, \bibinfo {author} {\bibfnamefont
  {K.}~\bibnamefont {{Jooya}}}, \bibinfo {author} {\bibfnamefont
  {A.}~\bibnamefont {{Khodabandelou}}}, \bibinfo {author} {\bibfnamefont
  {N.}~\bibnamefont {{Ladizinsky}}}, \bibinfo {author} {\bibfnamefont
  {R.}~\bibnamefont {{Li}}}, \bibinfo {author} {\bibfnamefont {P.~A.}\
  \bibnamefont {{Lott}}}, \bibinfo {author} {\bibfnamefont {A.~J.~R.}\
  \bibnamefont {{MacDonald}}}, \bibinfo {author} {\bibfnamefont
  {D.}~\bibnamefont {{Marsden}}}, \bibinfo {author} {\bibfnamefont
  {G.}~\bibnamefont {{Marsden}}}, \bibinfo {author} {\bibfnamefont
  {T.}~\bibnamefont {{Medina}}}, \bibinfo {author} {\bibfnamefont
  {R.}~\bibnamefont {{Molavi}}}, \bibinfo {author} {\bibfnamefont
  {R.}~\bibnamefont {{Neufeld}}}, \bibinfo {author} {\bibfnamefont
  {M.}~\bibnamefont {{Norouzpour}}}, \bibinfo {author} {\bibfnamefont
  {T.}~\bibnamefont {{Oh}}}, \bibinfo {author} {\bibfnamefont {I.}~\bibnamefont
  {{Pavlov}}}, \bibinfo {author} {\bibfnamefont {I.}~\bibnamefont
  {{Perminov}}}, \bibinfo {author} {\bibfnamefont {T.}~\bibnamefont
  {{Prescott}}}, \bibinfo {author} {\bibfnamefont {C.}~\bibnamefont {{Rich}}},
  \bibinfo {author} {\bibfnamefont {Y.}~\bibnamefont {{Sato}}}, \bibinfo
  {author} {\bibfnamefont {B.}~\bibnamefont {{Sheldan}}}, \bibinfo {author}
  {\bibfnamefont {G.}~\bibnamefont {{Sterling}}}, \bibinfo {author}
  {\bibfnamefont {L.~J.}\ \bibnamefont {{Swenson}}}, \bibinfo {author}
  {\bibfnamefont {N.}~\bibnamefont {{Tsai}}}, \bibinfo {author} {\bibfnamefont
  {M.~H.}\ \bibnamefont {{Volkmann}}}, \bibinfo {author} {\bibfnamefont
  {J.~D.}\ \bibnamefont {{Whittaker}}}, \bibinfo {author} {\bibfnamefont
  {W.}~\bibnamefont {{Wilkinson}}}, \bibinfo {author} {\bibfnamefont
  {J.}~\bibnamefont {{Yao}}}, \bibinfo {author} {\bibfnamefont
  {H.}~\bibnamefont {{Neven}}}, \bibinfo {author} {\bibfnamefont {J.~P.}\
  \bibnamefont {{Hilton}}}, \bibinfo {author} {\bibfnamefont {E.}~\bibnamefont
  {{Ladizinsky}}}, \bibinfo {author} {\bibfnamefont {M.~W.}\ \bibnamefont
  {{Johnson}}},\ and\ \bibinfo {author} {\bibfnamefont {M.~H.}\ \bibnamefont
  {{Amin}}},\ }\href@noop {} {\bibfield  {journal} {\bibinfo  {journal} {arXiv
  e-prints}\ ,\ \bibinfo {eid} {arXiv:1911.03446}} (\bibinfo {year} {2019})},\
  \Eprint {https://arxiv.org/abs/1911.03446} {arXiv:1911.03446 [quant-ph]}
  \BibitemShut {NoStop}%
\bibitem [{\citenamefont {Sylju{\aa}sen}\ and\ \citenamefont
  {Sandvik}(2002)}]{OFS2002}%
  \BibitemOpen
  \bibfield  {author} {\bibinfo {author} {\bibfnamefont {O.~F.}\ \bibnamefont
  {Sylju{\aa}sen}}\ and\ \bibinfo {author} {\bibfnamefont {A.~W.}\ \bibnamefont
  {Sandvik}},\ }\href {http://prola.aps.org/abstract/PRE/v66/i4/e046701}
  {\bibfield  {journal} {\bibinfo  {journal} {Physical Review E}\ }\textbf
  {\bibinfo {volume} {66}},\ \bibinfo {pages} {046701} (\bibinfo {year}
  {2002})}\BibitemShut {NoStop}%
\bibitem [{\citenamefont {Yan}\ \emph {et~al.}(2019)\citenamefont {Yan},
  \citenamefont {Wu}, \citenamefont {Liu}, \citenamefont {Sylju{\aa}sen},
  \citenamefont {Lou},\ and\ \citenamefont {Chen}}]{yan2019sweeping}%
  \BibitemOpen
  \bibfield  {author} {\bibinfo {author} {\bibfnamefont {Z.}~\bibnamefont
  {Yan}}, \bibinfo {author} {\bibfnamefont {Y.}~\bibnamefont {Wu}}, \bibinfo
  {author} {\bibfnamefont {C.}~\bibnamefont {Liu}}, \bibinfo {author}
  {\bibfnamefont {O.~F.}\ \bibnamefont {Sylju{\aa}sen}}, \bibinfo {author}
  {\bibfnamefont {J.}~\bibnamefont {Lou}},\ and\ \bibinfo {author}
  {\bibfnamefont {Y.}~\bibnamefont {Chen}},\ }\href
  {https://journals.aps.org/prb/abstract/10.1103/PhysRevB.99.165135} {\bibfield
   {journal} {\bibinfo  {journal} {Physical Review B}\ }\textbf {\bibinfo
  {volume} {99}},\ \bibinfo {pages} {165135} (\bibinfo {year}
  {2019})}\BibitemShut {NoStop}%
\bibitem [{\citenamefont {Yan}(2020)}]{yan2020improved}%
  \BibitemOpen
  \bibfield  {author} {\bibinfo {author} {\bibfnamefont {Z.}~\bibnamefont
  {Yan}},\ }\href {https://arxiv.org/abs/2011.08457} {\bibfield  {journal}
  {\bibinfo  {journal} {arXiv preprint arXiv:2011.08457}\ } (\bibinfo {year}
  {2020})}\BibitemShut {NoStop}%
\bibitem [{\citenamefont {Bl\"ote}\ and\ \citenamefont
  {Swendsen}(1979)}]{Bloete1979}%
  \BibitemOpen
  \bibfield  {author} {\bibinfo {author} {\bibfnamefont {H.~W.~J.}\
  \bibnamefont {Bl\"ote}}\ and\ \bibinfo {author} {\bibfnamefont {R.~H.}\
  \bibnamefont {Swendsen}},\ }\href
  {https://doi.org/10.1103/PhysRevLett.43.799} {\bibfield  {journal} {\bibinfo
  {journal} {Phys. Rev. Lett.}\ }\textbf {\bibinfo {volume} {43}},\ \bibinfo
  {pages} {799} (\bibinfo {year} {1979})}\BibitemShut {NoStop}%
\bibitem [{\citenamefont {Domany}\ \emph {et~al.}(1978)\citenamefont {Domany},
  \citenamefont {Schick}, \citenamefont {Walker},\ and\ \citenamefont
  {Griffiths}}]{Domany1978}%
  \BibitemOpen
  \bibfield  {author} {\bibinfo {author} {\bibfnamefont {E.}~\bibnamefont
  {Domany}}, \bibinfo {author} {\bibfnamefont {M.}~\bibnamefont {Schick}},
  \bibinfo {author} {\bibfnamefont {J.~S.}\ \bibnamefont {Walker}},\ and\
  \bibinfo {author} {\bibfnamefont {R.~B.}\ \bibnamefont {Griffiths}},\ }\href
  {https://doi.org/10.1103/PhysRevB.18.2209} {\bibfield  {journal} {\bibinfo
  {journal} {Phys. Rev. B}\ }\textbf {\bibinfo {volume} {18}},\ \bibinfo
  {pages} {2209} (\bibinfo {year} {1978})}\BibitemShut {NoStop}%
\bibitem [{\citenamefont {Domany}\ and\ \citenamefont
  {Schick}(1979)}]{Domany1979}%
  \BibitemOpen
  \bibfield  {author} {\bibinfo {author} {\bibfnamefont {E.}~\bibnamefont
  {Domany}}\ and\ \bibinfo {author} {\bibfnamefont {M.}~\bibnamefont
  {Schick}},\ }\href {https://doi.org/10.1103/PhysRevB.20.3828} {\bibfield
  {journal} {\bibinfo  {journal} {Phys. Rev. B}\ }\textbf {\bibinfo {volume}
  {20}},\ \bibinfo {pages} {3828} (\bibinfo {year} {1979})}\BibitemShut
  {NoStop}%
\bibitem [{\citenamefont {Alexander}(1975)}]{Alexander1975}%
  \BibitemOpen
  \bibfield  {author} {\bibinfo {author} {\bibfnamefont {S.}~\bibnamefont
  {Alexander}},\ }\href
  {https://doi.org/https://doi.org/10.1016/0375-9601(75)90766-5} {\bibfield
  {journal} {\bibinfo  {journal} {Physics Letters A}\ }\textbf {\bibinfo
  {volume} {54}},\ \bibinfo {pages} {353 } (\bibinfo {year}
  {1975})}\BibitemShut {NoStop}%
\bibitem [{\citenamefont {Damle}(2015)}]{Damle2015}%
  \BibitemOpen
  \bibfield  {author} {\bibinfo {author} {\bibfnamefont {K.}~\bibnamefont
  {Damle}},\ }\href {https://doi.org/10.1103/PhysRevLett.115.127204} {\bibfield
   {journal} {\bibinfo  {journal} {Phys. Rev. Lett.}\ }\textbf {\bibinfo
  {volume} {115}},\ \bibinfo {pages} {127204} (\bibinfo {year}
  {2015})}\BibitemShut {NoStop}%
\bibitem [{\citenamefont {Biswas}\ and\ \citenamefont
  {Damle}(2018)}]{Biswas2018}%
  \BibitemOpen
  \bibfield  {author} {\bibinfo {author} {\bibfnamefont {S.}~\bibnamefont
  {Biswas}}\ and\ \bibinfo {author} {\bibfnamefont {K.}~\bibnamefont {Damle}},\
  }\href {https://doi.org/10.1103/PhysRevB.97.085114} {\bibfield  {journal}
  {\bibinfo  {journal} {Phys. Rev. B}\ }\textbf {\bibinfo {volume} {97}},\
  \bibinfo {pages} {085114} (\bibinfo {year} {2018})}\BibitemShut {NoStop}%
\bibitem [{\citenamefont {José}\ \emph {et~al.}(1977)\citenamefont {José},
  \citenamefont {Kadanoff}, \citenamefont {Kirkpatrick},\ and\ \citenamefont
  {Nelson}}]{Jose1977}%
  \BibitemOpen
  \bibfield  {author} {\bibinfo {author} {\bibfnamefont {J.~V.}\ \bibnamefont
  {José}}, \bibinfo {author} {\bibfnamefont {L.~P.}\ \bibnamefont {Kadanoff}},
  \bibinfo {author} {\bibfnamefont {S.}~\bibnamefont {Kirkpatrick}},\ and\
  \bibinfo {author} {\bibfnamefont {D.~R.}\ \bibnamefont {Nelson}},\ }\href
  {https://doi.org/10.1103/PhysRevB.16.1217} {\bibfield  {journal} {\bibinfo
  {journal} {Physical Review B}\ }\textbf {\bibinfo {volume} {16}},\ \bibinfo
  {pages} {1217} (\bibinfo {year} {1977})}\BibitemShut {NoStop}%
\bibitem [{\citenamefont {Challa}\ and\ \citenamefont
  {Landau}(1986)}]{Challa1986}%
  \BibitemOpen
  \bibfield  {author} {\bibinfo {author} {\bibfnamefont {M.~S.~S.}\
  \bibnamefont {Challa}}\ and\ \bibinfo {author} {\bibfnamefont {D.~P.}\
  \bibnamefont {Landau}},\ }\href {https://doi.org/10.1103/PhysRevB.33.437}
  {\bibfield  {journal} {\bibinfo  {journal} {Phys. Rev. B}\ }\textbf {\bibinfo
  {volume} {33}},\ \bibinfo {pages} {437} (\bibinfo {year} {1986})}\BibitemShut
  {NoStop}%
\bibitem [{\citenamefont {Sachdev}(2011)}]{Sachdev}%
  \BibitemOpen
  \bibfield  {author} {\bibinfo {author} {\bibfnamefont {S.}~\bibnamefont
  {Sachdev}},\ }\href {https://doi.org/DOI: 10.1017/CBO9780511973765} {\emph
  {\bibinfo {title} {Quantum Phase Transitions}}},\ \bibinfo {edition} {2nd}\
  ed.\ (\bibinfo  {publisher} {Cambridge University Press},\ \bibinfo {address}
  {Cambridge},\ \bibinfo {year} {2011})\BibitemShut {NoStop}%
\bibitem [{\citenamefont {Fradkin}(2013)}]{Fradkin}%
  \BibitemOpen
  \bibfield  {author} {\bibinfo {author} {\bibfnamefont {E.}~\bibnamefont
  {Fradkin}},\ }\href
  {https://www.cambridge.org/core/books/field-theories-of-condensed-matter-physics/EABECB65A0F4F9289B2737A6DD3E6C0D}
  {\emph {\bibinfo {title} {Field theories of condensed matter physics}}}\
  (\bibinfo  {publisher} {Cambridge University Press},\ \bibinfo {year}
  {2013})\ \bibinfo {type} {Book}~\bibinfo {chapter} {4}, pp.\ \bibinfo {pages}
  {63--89}\BibitemShut {NoStop}%
\bibitem [{\citenamefont {Nagaosa}(1999)}]{NagaosaSCES}%
  \BibitemOpen
  \bibfield  {author} {\bibinfo {author} {\bibfnamefont {N.}~\bibnamefont
  {Nagaosa}},\ }\href {https://www.springer.com/gp/book/9783540659815} {\emph
  {\bibinfo {title} {Quantum field theory in strongly correlated electronic
  systems}}}\ (\bibinfo  {publisher} {Springer Science \& Business Media},\
  \bibinfo {year} {1999})\BibitemShut {NoStop}%
\bibitem [{\citenamefont {Francesco}\ \emph {et~al.}(2012)\citenamefont
  {Francesco}, \citenamefont {Mathieu},\ and\ \citenamefont
  {S{\'e}n{\'e}chal}}]{FrancescoCFT}%
  \BibitemOpen
  \bibfield  {author} {\bibinfo {author} {\bibfnamefont {P.}~\bibnamefont
  {Francesco}}, \bibinfo {author} {\bibfnamefont {P.}~\bibnamefont {Mathieu}},\
  and\ \bibinfo {author} {\bibfnamefont {D.}~\bibnamefont {S{\'e}n{\'e}chal}},\
  }\href@noop {} {\emph {\bibinfo {title} {Conformal field theory}}}\ (\bibinfo
   {publisher} {Springer Science \& Business Media},\ \bibinfo {year}
  {2012})\BibitemShut {NoStop}%
\bibitem [{\citenamefont {Henkel}(2013)}]{HenkelCFT}%
  \BibitemOpen
  \bibfield  {author} {\bibinfo {author} {\bibfnamefont {M.}~\bibnamefont
  {Henkel}},\ }\href@noop {} {\emph {\bibinfo {title} {Conformal invariance and
  critical phenomena}}}\ (\bibinfo  {publisher} {Springer Science \& Business
  Media},\ \bibinfo {year} {2013})\BibitemShut {NoStop}%
\bibitem [{Note1()}]{Note1}%
  \BibitemOpen
  \bibinfo {note} {$h\rightarrow h \Lambda ^{3/2}$ makes $h$ a dimensionless
  quantity so that it is comparable with unity, and all other terms are already
  marginal.}\BibitemShut {Stop}%
\bibitem [{Note2()}]{Note2}%
  \BibitemOpen
  \bibinfo {note} {The second structure constant is valid only when $n,m\neq
  0$, for only vortex operator calculation involved.}\BibitemShut {Stop}%
\end{thebibliography}%

\end{document}